\documentclass[prl,twocolumn,aps,amsmath,amssymb,superscriptaddress,showpacs]{revtex4-2}

\usepackage{amsmath,amssymb}
\usepackage{graphicx}
\usepackage{appendix}
\usepackage{subcaption}
\usepackage[usenames,dvipsnames]{color}
\usepackage[colorlinks=true, citecolor=blue, linkcolor=blue, urlcolor=blue]{hyperref}
\usepackage{hyphenat}
\usepackage{txfonts}
\usepackage{dsfont}
\usepackage{enumitem}
\usepackage[utf8]{inputenc}

\begin{document}

\title{Theory of Quantum-Enhanced Stimulated Raman Scattering}
\author{Frank Schlawin}
\affiliation{Max Planck Institute for the Structure and Dynamics of Matter, Luruper Chaussee 149, 22761 Hamburg, Germany}
\affiliation{University of Hamburg, Luruper Chaussee 149, Hamburg, Germany}
\affiliation{The Hamburg Centre for Ultrafast Imaging, Luruper Chaussee 149, Hamburg D-22761, Germany}
\author{Manuel Gessner}
\affiliation{Departament de F\'isica Te\`orica and IFIC, Universitat de Val\`encia-CSIC, C/Dr Moliner 50, 46100 Burjassot (Valencia), Spain}

\date{November 2023}

\begin{abstract}
Stimulated Raman scattering (SRS) is a powerful method for label‐free imaging and spectroscopy of materials. 
Recent experiments have shown that quantum-enhanced Raman scattering can surpass the shot noise limit and improve the sensitivity substantially. Here, we introduce a full theory of quantum-enhanced SRS based on the framework of quantum metrology. Our results enable the assessment of quantum-enhancements of arbitrary measurement strategies and identify optimal measurement observables that extract maximal information about the signal. We use this to identify the optimal employment of squeezed states in SRS, highlighting the potential to improve quantum gains beyond those observed in recent experiments. Our work establishes the theoretical foundation for understanding and approaching the quantum limits of precision in SRS, and provide a tool to discuss nonlinear spectroscopy and imaging more broadly.
\end{abstract}

\maketitle

\textit{Introduction.---}Raman scattering is a nonlinear wave-mixing process, whereby a photon inelastically scatters from a sample, while creating or annihilating a phonon. Its importance and ubiquity in physical, chemical and even biological research is difficult to overstate~\cite{Evans2008, Cheng2015, VoronineZhangSokolovScully+2018+523+548}. 
Since the Raman process directly probes molecular bonds, it enables the label-free acquisition of compositional and functional information of chemical constituents of biological systems~\cite{CampJr2015, Prince2017, Tipping2024}. 
Time-resolved Raman spectroscopy provides insights into ultrafast structural dynamics and enables the direct monitoring of vibrational wavepackets in molecules~\cite{Kukura2007, Batignani2024} or quantum materials~\cite{Pellatz2021, Chou2024}. 
However, spontaneous Raman cross sections are typically extremely small, impeding the use of many Raman techniques in many applications. This is why the recent decades have seen the rapid development and maturing of stimulated and coherent Raman techniques, including notably stimulated Raman scattering (SRS) and coherent anti-Stokes Raman scattering (CARS), as well as countless closely related techniques~\cite{Rahav2010, Raman-book}. 
Raman-based imaging and spectroscopy technology has reached a stage where it is limited by the laser shot noise or the presence of undesired non-resonant background signals. 

One particularly promising area of research, where further advances may be imagined, lies in the use of quantum-enhanced measurement strategies that are designed within the framework of quantum metrology~\cite{Bowen-review}.  Uniting parameter estimation theory with quantum measurements, this framework establishes fundamental precision limits in diverse settings~\cite{VittorioNP2011,Pezze2014}. Systematic optimization over measurement observables can enhance sensitivity; in some cases even when the probe state is classical.
This can lead to surprising insights, such as the possibility to overcome the Rayleigh limit in superresolution imaging based on optimal measurements, even when using thermal light sources~\cite{PhysRevX.6.031033, TsangReview}. When nonclassical probe states are accessible, the limits of measurement precision can be pushed even further by engineering squeezed or entangled states with reduced quantum fluctuations~\cite{Caves1981,Pezze2014}, as prominently realized in gravitational wave detectors~\cite{VirgoPRL2019, TsePRL2019}.

Theoretical studies have highlighted the promising potential of using entangled photons in Raman measurements, suggesting that they can help reduce the nonresonant background signal and improve the scaling of nonlinear signals~\cite{Dorfman2014, MUNKHBAATAR2017581, Chen2021, Schlawin2021, Fan2024, Jadoun2024}. Similarly, the use of squeezed light sources, which can reduce photon number fluctuations at the optimal operating point of classical Raman experiments, points to exciting opportunities. Proof-of-principle experiments have already demonstrated that quantum-enhanced Raman measurements are within reach, whether with narrowband fields~\cite{deAndrade:20} or ultrafast pulses~\cite{Casacio2021, Xu22, Terrasson24}, and in stimulated Brillouin scattering~\cite{Li2022, Li2024}. These advances, along with parallel progress in stimulated emission imaging~\cite{Triginer-Garces2020}, highlight the role of squeezed light in near-future nonlinear measurement setups. The observed advantages could be further amplified through postprocessing techniques~\cite{Gong23} or the use of nonlinear interferometers~\cite{Michael2019}. The enhancement of detection efficiency via photon-counting in spontaneous Raman spectroscopy~\cite{Li24}, and its potential combination with nanoplasmonic sensing~\cite{XavierYuJonesZossimovaVollmer+2021+1387+1435}, add another promising dimension. Recent initial efforts to address spectroscopic challenges with quantum metrology approaches~\cite{Gianani2021, SanchezMunoz2021, Panahiyan2023, Albarelli2023, Karsa2024} further hint at a rich landscape of possibilities. Together, these developments serve as compelling indicators of a promising direction in quantum-enhanced Raman spectroscopy that calls for the development of a comprehensive quantum theoretical framework and a systematic exploration of its precision limits.

In this manuscript, we establish such a quantum metrological theory of SRS. To this end, we derive a fully quantum description of nonlinear light-matter interactions by formulating them as a scattering theory that connects the input and output states of light and carrying out a cumulant expansion of the light-matter interaction. Moreover, we relate our scattering description to the established double-sided Feynman diagrams commonly used in nonlinear spectroscopy to simulate optical signals. As a result, conventional molecular or many-body dynamics simulations used for calculating these signals can be readily adapted to analyze quantum-enhanced Raman scattering. Using the tools of quantum metrology, we then quantify the potential quantum advantage in SRS. 
We use this general framework to assess the performance of coherent (i.e. classical) states, squeezed coherent  probe pulses (as experimentally implemented in Refs.~\cite{deAndrade:20, Casacio2021}), and two-mode squeezed states. 
While in all cases the sensitivity increases quadratically with the mean photon number, our results reveal that sensitivity can be significantly increased by optimizing the distribution of photons over squeezed and coherent states and by using entangled probes.

\textit{Quantum theory of nonlinear light-matter interactions.---}We consider a sample system interacting with one or several quantum state(s) of light. 
Before their interaction, sample and light are in a separable state, $\varrho_0 = \varrho_{s} \otimes \varrho_{in}$, where $\varrho_s$ is the sample density matrix and $\varrho_{in}$ the photonic one. 
The formally exact combined light-matter density matrix at a time $t$ is given by Dyson series in the interaction picture with respect to the molecular and the field Hamiltonians, 
\begin{align}
    \varrho (t) &= \mathcal{T} \exp \left[ - \frac{i}{\hbar} \int^t_{-\infty} \!\! d\tau \hat{H}_{I, -} (\tau) \right] \varrho_0. \label{eq:rho(t)}
\end{align}
Here, $\hat{H}_I$ is the interaction Hamiltonian, $\mathcal{T}$ the time-ordering operator, and the minus subscript indicates the commutator Liouville space operators, with $\hat{H}_{I, -} \hat{X} \equiv \hat{H}_I \hat{X} - \hat{X} \hat{H}_I$.

We will be concerned with measurements on the output light field, i.e. on the field density matrix long after it has passed through the sample in the limit $t\rightarrow \infty$.
As we only carry out measurements on the light fields, we trace out the molecular degrees of freedom, tr$_s \{ \ldots \}$, and obtain a quantum map $\Phi$ for the photonic degrees of freedom (see SM for the explicit form of $\Phi$),
\begin{align}
    \varrho_{out} &= \Phi [\varrho_{in}]. \label{eq.Phi-def}
\end{align}
We next carry out a cumulant expansion in the light-matter interaction, assuming that the interaction Hamiltonian scales as a small expansion parameter, $\hat{H}_I \sim\sqrt{\Gamma}$, i.e. 
we approximate the quantum map $\Phi$ by an exponential map, 
\begin{align} \label{eq.Phi}
    \Phi &\simeq \exp \left[ \sum_{n= 1}^M \Gamma^{n/2} \mathcal{K}_n  \right],
\end{align}
where the expansion is terminated at a finite order $M$. 
The connected correlation functions of this expansion are given by time-ordered correlation functions of the form
\begin{align}
    \left( - \frac{i}{\hbar} \right)^n \int^{\infty}_{-\infty} \!\! d\tau_n \ldots \int^{\tau_2}_{-\infty} \!\! d\tau_1 \; \text{tr}_{s} \left\{ \hat{H}_{I, -} (\tau_n) \ldots \hat{H}_{I, -} (\tau_1) \varrho_{s} \right\}.
\end{align}
These expressions may be evaluated with the usual Feynman diagram methods~\cite{Shaul_book}. In this case, however, since only the sample degrees of freedom are traced out, and the interaction Hamiltonian acts on both Hilbert spaces, the result is a superoperator acting on the photonic degrees of freedom.
Once we determine $\Phi$ in this manner, this formalism may be used to analyze, in principle, any nonlinear wave-mixing signal from a quantum metrology point of view. 
We stress, however, that a cumulant expansion is not guaranteed to produce a CTPT map, as is the case in the derivation of time-convolutionless master equations~\cite{BreuerPetruccione}

\begin{figure}
    \centering
    \includegraphics[clip, trim={0 0 340 0}, width=0.49\textwidth]{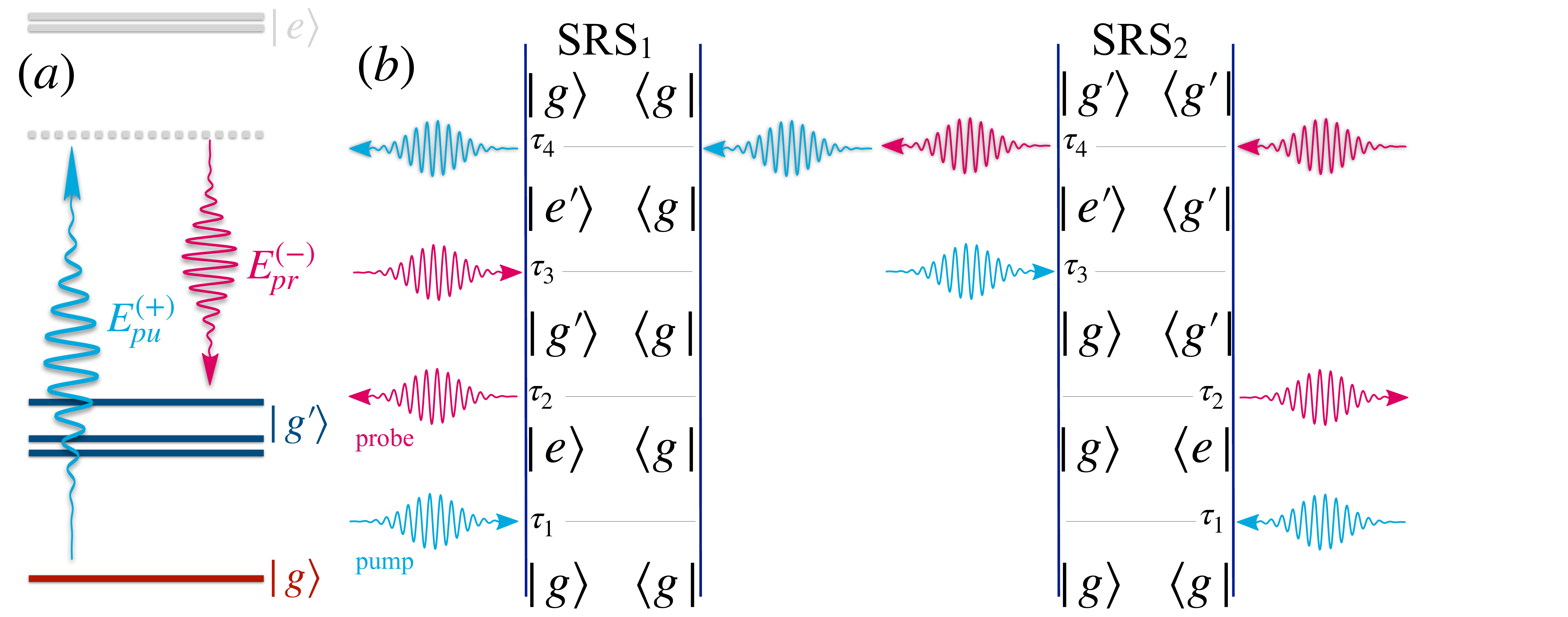}
    \caption{
    Level schemes and signal diagrams considered in this paper.
    (a) The Raman transition between ground state $\vert g \rangle$ and vibrationally excited state $\vert g ' \rangle$ takes place via a far off-resonant electronic state $\vert e \rangle$. The bandwidths of the Raman pump and probe pulses, $E_{pu}$ and $E_{pr}$, are considered smaller than the transition frequency $\omega_{g'} - \omega_g$.  (b) The Raman signal is generated by two double-sided Feynman diagrams representing the parametric contribution SRS$_1$ and the dissipative contribution SRS$_2$, respectively.
    } 
    \label{fig:setup}
\end{figure}

\textit{Stimulated Raman scattering.---}Here, we apply the above theory to SRS, driven by two broadband modes, i.e. we write the electric field operators as the sum of pump and probe pulses $\hat{E}(t) = \hat{E}_{pu} (t) + \hat{E}_{pr} (t)$, described by the operators
\begin{align}
    \hat{E}_{pu} (t) &= \epsilon_{pu} \psi_{pu} (t) \hat{A}, \label{eq.E_pu} \\
    \hat{E}_{pr} (t) &= \epsilon_{pr} \psi_{pr} (t) \hat{B}, \label{eq.E_pr}
\end{align}
where $\epsilon_{pu/pr}$ is the vacuum field strength at the sample position (see SM), 
$\psi_{pu/pr}$ are the respective normalized time domain mode functions, and $\hat{A} = \int dt \; \psi^\ast_{pu} (t) \hat{a}_{pu}(t)$, and 
$\hat{B} = \int dt \; \psi^\ast_{pr} (t) \hat{a}_{pr} (t)$
are broadband operators which preserve their bosonic commutation relations by virtue of the normalization of the mode functions~\cite{Brecht2015}.

\begin{figure*}
    \centering
    \includegraphics[clip, trim={0 740 800 0},width=.9\textwidth]{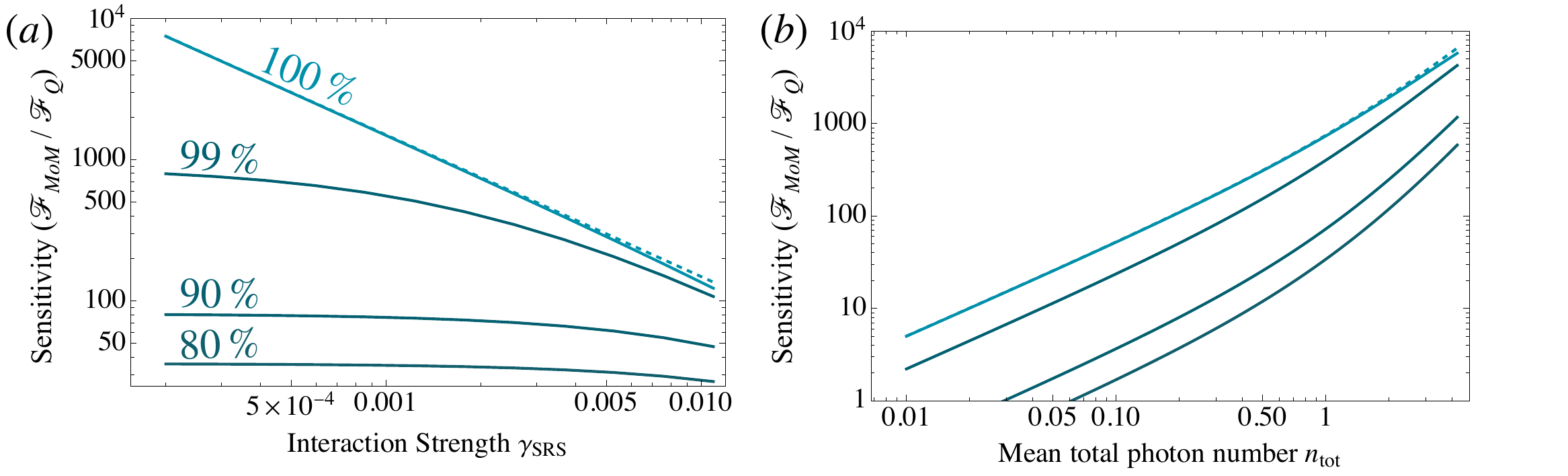}
    \caption{Metrological performance of two-mode squeezed state for SRS measurements: (a) Quantum Fisher Information $\mathcal{F}_Q$ (dashed) and method of moments information $\mathcal{F}_{MoM} (\Delta \hat{n})$ are plotted vs. interaction strength. Darker plots correspond to $\mathcal{F}_{MoM}$ with imperfect photodetectors with indicated quantum efficiency (see SM). The total photon number is fixed at $n_{tot} = 1$. (b) same as (a) vs. mean total photon number with $\gamma_{SRS} = 2 \times 10^{-3}$.}
    \label{fig:tms-plot}
\end{figure*}

We consider the SRS signal of $N$ molecules, each described by the model illustrated in Fig.~\ref{fig:setup}(a). It involves one or several high-frequency vibrational states $\vert g'\rangle$ which are two-photon coupled to the ground state $\vert g \rangle$. Raman transitions involving pump and probe pulses can excite the system and transfer energy from the pump to the probe beam (Raman gain). 
We neglect the inverse process (Raman loss) since thermal occupation of high-frequency phonons is small (though it could be amended straightforwardly). 
Linear photon losses can be neglected as well, as long as high-energy excited states $\vert e\rangle$ are far detuned from the photonic frequencies involved. 
Only the fourth-order contribution $\mathcal{K}_4$ becomes relevant. 
Following the established approach in laser spectroscopy, we evaluate the contributions with the help of double-sided Feynman diagrams. The diagrams contributing to the SRS signal are shown in Fig.~\ref{fig:setup}(b). Using the Feynman diagram rules (see SM), they translate to 
\begin{align} \label{eq.K_SRS}
    \mathcal{K}_{SRS} [\varrho] &= - i h_{SRS} \left[ \hat{L}^\dagger \hat{L}, \varrho \right]
    +\gamma_{SRS} \left( \hat{L} \varrho \hat{L}^\dagger - \frac{1}{2} \{ \hat{L}^\dagger \hat{L}, \varrho \} \right),
\end{align}
where $\hat{L} = \hat{A} \hat{B}^\dagger$ is the operator describing the inelastic scattering of a pump photon into the probe beam.  
We have further defined
\begin{align}
    h_{SRS} &= N \epsilon_{pu}^2 \epsilon_{pr}^2 \sum_{g'} | \alpha_{g' g} |^2  \int \frac{d\omega_-}{2\pi} \frac{\omega_- - \omega_{g'g}}{ (\omega_- - \omega_{g'g})^2 + \gamma_{g'g}^2 } \vert \Phi (\omega_-) \vert^2, \label{eq.h_SRS} \\
    \gamma_{SRS} &= 2 N\epsilon_{pu}^2 \epsilon_{pr}^2 \sum_{g'} | \alpha_{g' g} |^2 \int \frac{d\omega_-}{2\pi} \frac{\gamma_{g'g}}{ (\omega_- - \omega_{g'g})^2 + \gamma_{g'g}^2 } \vert \Phi (\omega_-) \vert^2, \label{eq.gamma_SRS}
\end{align}
with the two-photon spectral density~\cite{Mukamel2009} at frequency $\omega$, $\Phi (\omega) = \int\frac{d\omega'}{2\pi} \psi_{pu} (\omega') \psi^\ast_{pr} (\omega' - \omega)$, and $\alpha_{g'g} = \mu_{ge} \mu_{g' e}^\ast / (\hbar^2 \Delta)$ is the Raman polarizability with $\Delta$ the detuning from the excited state $\vert e \rangle$. Although $h_{SRS}$ and $\gamma_{SRS}$ seem to describe a frequency and a decay rate, respectively, they are in fact dimensionless quantities in Eq.~(\ref{eq.K_SRS}).  
The result~(\ref{eq.K_SRS}), which generalizes earlier work from the 1970's~\cite{Walls1970,Walls1971} to pulsed excitation, neglects intermolecular coherence and propagation effects in macroscopic samples that may lead to new effects~\cite{Raymer1980, Raymer1985}. 
As we show in the SM, the semiclassical limit may be recovered from this form, and the formalism also allows to account for nonresonant background contributions, if necessary. 

In the following, we will focus on the estimation of $\gamma_{SRS}$, for the case of a single high-frequency Photon non mode. This is an appropriate model for, e.g., polystyrene used in~\cite{Casacio2021}. 
When the pump and probe fields are on two-photon resonance with the vibration, i.e. $\omega_{pu} - \omega_{pr} = \omega_{g'g}$, the Hamiltonian contribution vanishes, $h_{SRS} = 0$, and the dynamics is purely incoherent with $\gamma_{SRS} \neq 0$, such that $\gamma_{SRS}$ determines the Raman signal.

\textit{Quantum metrology.---}Equipped with the full quantum description of the SRS process, we are now in a position to identify the quantum limit of precision for estimations of the parameter of interest $\gamma_{SRS}$. This limit is given by the quantum Cramér-Rao bound (QCRB)~\cite{Pezze2014},
\begin{align}
    (\Delta\gamma_{SRS, \mathrm{est}})^2\geq \frac{1}{m \mathcal{F}_Q[\rho(\gamma_{SRS})]},
\end{align}
that sets a lower limit on the variance of any unbiased estimator $\gamma_{SRS, \mathrm{est}}$ of the Raman strength $\gamma_{SRS}$, with $\mathcal{F}_Q[\rho(\gamma_{SRS})]$ denoting the quantum Fisher information (QFI) that quantifies the state $\rho(\gamma_{SRS})$’s sensitivity to small changes in $\gamma_{SRS}$~\cite{BraunsteinPRL1994}, and $m$ representing the number of repeated measurements. This bound is saturated by an optimal estimator, such as the maximum likelihood estimator in the asymptotic limit of many measurements, $m\gg 1$, applied to the data obtained from an optimal observable. This observable can be theoretically determined, but in some situations it may be inaccessible for the experiment in question.

A simple, practical estimator can be constructed from the expectation value of any specific accessible observable $X$ using the method of moments (MoM). After previous calibration of the behavior of $\langle X\rangle_{\rho(\gamma_{SRS})}=\mathrm{Tr}\{X\rho(\gamma_{SRS})\}$ as a function of $\gamma_{SRS}$, a sample average of measurements of $X$ can be used to estimate the value of $\gamma_{SRS}$, which, in the asymptotic limit, $m\gg 1$, yields the sensitivity
\begin{align} \label{eq:MoM}
    (\Delta\gamma_{SRS, \mathrm{est}})^2=\frac{1}{m\mathcal{F}_{MoM} [X, \rho(\gamma_{SRS})]},
\end{align}
where $\mathcal{F}_{MoM} = | \partial \langle X\rangle_{\rho(\gamma_{SRS})} / \partial\gamma_{SRS}|^2 / (\Delta X)_{\rho(\gamma_{SRS})}^2$. 
When the observable $X$ is chosen optimally, the above sensitivity matches that of the QFI~\cite{gessner_metrological_2019}; otherwise, the MoM still offers a lower bound, $\mathcal{F}_{MoM} [X, \rho(\gamma_{SRS})]\leq\mathcal{F}_Q[\rho(\gamma_{SRS})]$, obtained from a practical and flexible estimation strategy.

\textit{Two-mode squeezed state.---}We first demonstrate the potential of quantum metrology for SRS with an example where the QFI may be saturated in state-of-the-art experiments: 
Since the Lindblad operator in Eq.~(\ref{eq.K_SRS}) converts pump into probe photons, a quantum state with strong correlations between pump and probe should be very sensitive to even weak SRS events. 
Hence, a state such as a two-mode squeezed vacuum,
\begin{align}
    \vert \psi_{tms} \rangle &= \frac{1}{\cosh (r)}\sum_{n= 0}^\infty \tanh^n (r) \vert n \rangle_{pu} \otimes \vert n\rangle_{pr}, \label{eq:tms-state}
\end{align}
(here $r$ is the squeezing parameter)  
which shows perfect correlations between photons in pump and probe beam can be expected to be a highly sensitive probe for SRS scattering: 
The photon number difference $\Delta \hat{n} = \hat{n}_{pu} - \hat{n}_{pr}$ and its variance vanishes, Var$(\Delta \hat{n}) = 0$, and any difference thus stems from SRS events. 
This intuition is supported by our simulations in Fig.~\ref{fig:tms-plot}, where we numerically evaluate $\varrho_{out} = \exp [\mathcal{K}_{SRS}] \varrho_{in}$ and plot the QFI and MoM for measuring $\Delta \hat{n}$ as a function of the interaction strength $\gamma_{SRS}$. 
The MoM saturates the QFI (expect when $\gamma_{SRS}$ becomes very large) and is thus an optimal measurement for SRS detection with two-mode squeezed states. 
As a function of mean photon number [Fig.~\ref{fig:tms-plot}(b)], the QFI scales linearly when $\langle \hat{n}_{pu} +\hat{n}_{pr} \rangle \lesssim 1$, and quadratically when $\langle \hat{n}_{pu} +\hat{n}_{pr} \rangle > 1$. 
This behaviour is  reminiscent of other nonlinear signals such as two-photon absorption~\cite{SanchezMunoz2021, Panahiyan2023}. The linear scaling signifies the isolated pair regime, where the squeezed vacuum is composed of temporally separated pairs of entangled photons, 
where time-energy entanglement might allow for further metrological gains. 

\begin{figure}
    \centering
    \includegraphics[width=0.49\textwidth]{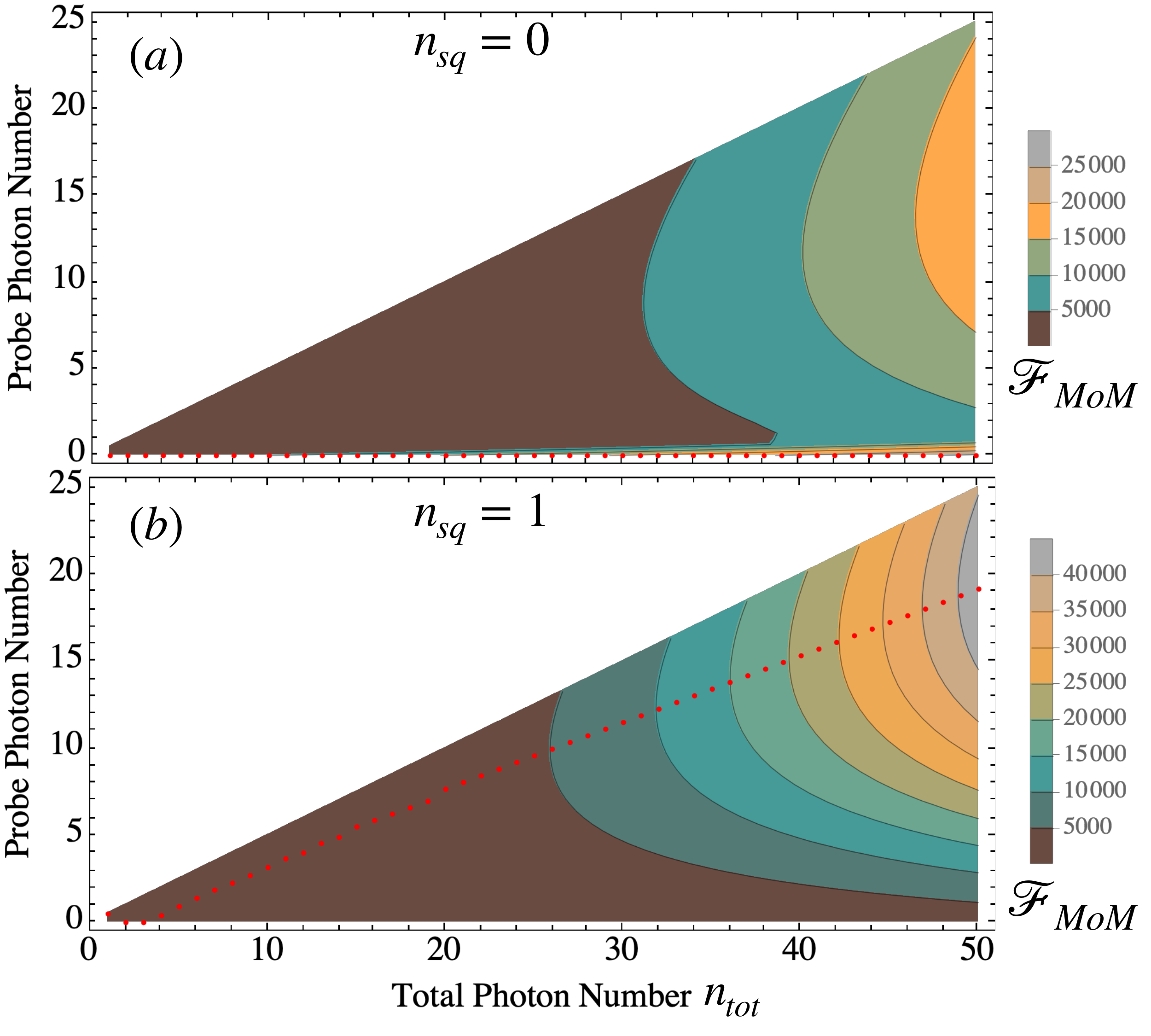}
    \caption{Optimization of squeezed probes: 
    $\mathcal{F}_{MoM}$, Eq.~(\ref{eq:MoM}), 
    is plotted vs. the mean total photon number $n_{tot}$ and the coherent probe photon number $\vert \alpha_{pr}\vert^2$ for different squeezing levels, $n_{sq} = \sinh^2 (r)$. 
    The relative phases are set such that a number squeezed probe state is generated, and we fix $\gamma_{SRS} = 2 \times 10^{-3}$. }
    \label{fig:example}
\end{figure}

It is worth noting that the QFI and MoM diverge when $\gamma_{SRS} \rightarrow 0$. In the case of the QFI, this divergence may be understood as a consequence of the incoherent nature of the SRS scattering [see Eq.~(\ref{eq.K_SRS})], as discussed e.g. in Ref.~\cite{Seveso_2020}. The change of the purity of the transmitted state causes the QFI to diverge. 
Perhaps more surprisingly, the MoM also diverges. This is, however, a direct consequence of the ideal correlations encapsulated in Eq.~(\ref{eq:tms-state}), and unstable against noise: 
If we consider, for instance, imperfect photodetection, the divergence immediately disappears [see Fig.~\ref{fig:tms-plot}(a)].

\textit{Possible metrological gain compared to recent experiments.---}Having demonstrated how to optimize SRS measurements with two-mode squeezed states, we compare their performance to the state of the art. In the following, we compare to optimized measurements with coherent pulses to derive the sensitivity achievable with classical resources, as well as with measurements with a squeezed coherent probe, as used in recent experiments~\cite{deAndrade:20,Casacio2021}. The full initial quantum state of light is given by
\begin{align}
    \vert \psi_{sq-coh} \rangle &= \vert \alpha_{pu} \rangle_{pu} \otimes \vert \alpha_{pr}, \zeta_{pr} \rangle_{pr}, \label{eq:sq-coh-state}
\end{align}
and to optimize metrological performance, we maximize the achievable sensitivity through variation of the parameters $\alpha_{pu}, \alpha_{pr}$ and $\zeta_{pr}$ for fixed total photon number $n_{tot} = \langle \hat{n}_{pu} + \hat{n}_{pr} \rangle$~\footnote{This target is motivated by potential photodamage or the molecules. Depending on experimental considerations, other optimization targets may also be imagined. }.

Cuts through this 5D optimization landscape (three amplitudes and two relative phases) are shown in Fig.~\ref{fig:example} for the optimization of $\mathcal{F}_{MoM}$.
When the total photon number is small, as is the case in our exact simulations, spontaneous Raman scattering is optimal and hence $\alpha_{pr}  = 0$~\footnote{We clarify here our model is insufficient to describe the spontaneous regime adequately, since only two modes are included in Eq.~(\ref{eq.K_SRS}). The 'true' spontaneous emission is isotropic, potentially necessitating a multimode description.}. 
When the mean total photon number crosses a critical value which scales as $n_{cr} \sim 1/ \gamma_{SRS}$ (as can be deduced from perturbation theory), a stimulated regime becomes optimal. This crossover can be pushed to lower photon numbers by the addition of squeezing [see Fig.~\ref{fig:example}(b)].

\begin{figure}
    \centering
    \includegraphics[width=0.45\textwidth]{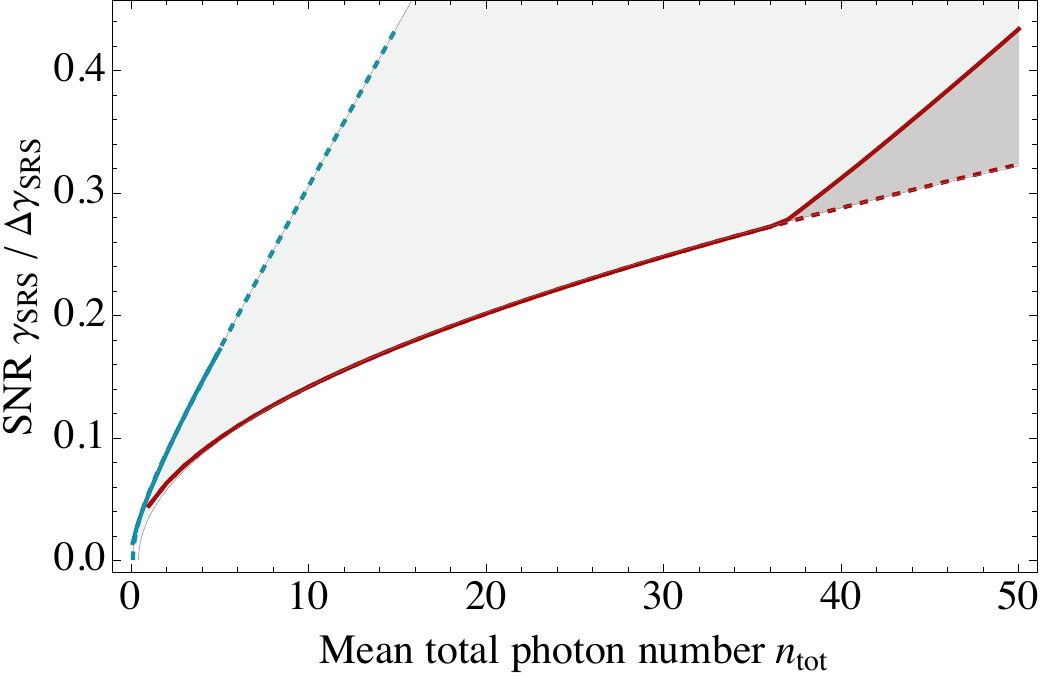}
    \caption{Metrological advantage with present day technology: 
    The asymptotic signal-to-noise ratio (SNR) per shot, $\gamma_{SRS} / \Delta\gamma_{SRS}$, according to Eq.~(\ref{eq:MoM}) with $m =1$
    with optimized squeezing (red, solid) and for optimized coherent states (red, dashed) are compared with $\mathcal{F}_{MoM} (\Delta\hat{n})$ for two-mode squeezed states (solid, blue) is plotted vs. the mean photon number $n_{tot}$. We fix $\gamma_{SRS} = 2 \times 10^{-3}$.
    }
    \label{fig:comparison}
\end{figure}

In Fig.~\ref{fig:comparison}, we summarize the metrological advantage that could be gained with existing technology: we compare the signal to noise ratio $\gamma_{SRS} / \Delta\gamma_{SRS}$ for a single experimental run (i.e. $m = 1$) for optimized coherent states and squeezed probes. The dark, shaded area shows the metrological advantage that can be achieved with squeezed probes (in the intensity regime here, the optimal coherent experiment is a spontaneous Raman experiment with $\alpha_{pr} = 0$. Eventually, the dashed line with also become quadratic in the stimulated regime).  Whereas a recent experiment~\cite{Casacio2021} reported a 20 \% increase of the signal to noise ratio, our simulations indicate that a much more substantial increase of the sensitivity is feasible even at moderate photon numbers, relying only on probe intensity measurements and optimally tuned squeezed probe states. But crucially, much larger gains (light shaded area) are possible through the use of optimized measurements, such as the photon number difference with two-mode squeezed states. 

\textit{Conclusions.---}We have developed a quantum metrological theory of stimulated Raman scattering, based on a cumulant expansion of the light-matter interaction. Our analysis of the SRS measurement with broadband pump and probe pulses demonstrates that substantial improvements of the achievable sensitivity beyond the recent experimental demonstrations are feasible with present-day technology. Our results further demonstrate that quantum correlations between pump and probe can push these enhancements per average photon number even further. Moreover, we identified a simple, metrologically optimal measurement scheme based on moment-based estimation and measurements of the photon number difference. 

Based on these results, further enhancements with multimode squeezed beams (i.e. exploiting time-energy entanglement and/or momentum entanglement) can be expected. Judging from the insights gained in the analysis of entanglement-enhanced two-photon absorption~\cite{Schlawin2017}, this enhancement will likely be sample-dependent via the response function and must be evaluated on case by case basis. Other interesting future developments of our work may emerge from an extension of our model to several phonon lines, where the necessity of estimating both $h_{SRS}$ and $\gamma_{SRS}$ complicates the analysis, but may also lead to different, potentially superior estimation strategies based on quantum multiparameter estimation~\cite{PhysRevLett.121.130503,Liu_2020,ALBARELLI2020126311,Gessner2020}. The theory presented here can be applied to analyze other nonlinear signals and serves as a foundation for systematic studies in quantum-enhanced nonlinear spectroscopy and imaging. 

\section{Acknowledgments}
FS acknowledges support from the Cluster of Excellence 'Advanced Imaging of Matter' of the Deutsche Forschungsgemeinschaft (DFG) - EXC 2056 - project ID 390715994. This work was supported by the project PID2023-152724NA-I00, with funding from MCIU/AEI/10.13039/501100011033 and FSE+. This work was funded by MCIN/AEI/10.13039/501100011033 and the European Union  ?NextGenerationEU? PRTR fund [RYC2021-031094-I], by the Ministry of Economic Affairs and Digital Transformation of the Spanish Government through the QUANTUM ENIA Project call?QUANTUM SPAIN Project, by the European Union through the Recovery, Transformation and Resilience Plan?NextGenerationEU within the framework of the Digital Spain 2026 Agenda, and by the CSIC Interdisciplinary Thematic Platform (PTI+) on Quantum Technologies (PTI-QTEP+). This work was supported through the project CEX2023-001292-S funded by MCIU/AEI.

\bibliography{bibliography_metrology}

\begin{thebibliography}{60}%
\makeatletter
\providecommand \@ifxundefined [1]{%
 \@ifx{#1\undefined}
}%
\providecommand \@ifnum [1]{%
 \ifnum #1\expandafter \@firstoftwo
 \else \expandafter \@secondoftwo
 \fi
}%
\providecommand \@ifx [1]{%
 \ifx #1\expandafter \@firstoftwo
 \else \expandafter \@secondoftwo
 \fi
}%
\providecommand \natexlab [1]{#1}%
\providecommand \enquote  [1]{``#1''}%
\providecommand \bibnamefont  [1]{#1}%
\providecommand \bibfnamefont [1]{#1}%
\providecommand \citenamefont [1]{#1}%
\providecommand \href@noop [0]{\@secondoftwo}%
\providecommand \href [0]{\begingroup \@sanitize@url \@href}%
\providecommand \@href[1]{\@@startlink{#1}\@@href}%
\providecommand \@@href[1]{\endgroup#1\@@endlink}%
\providecommand \@sanitize@url [0]{\catcode `\\12\catcode `\$12\catcode
  `\&12\catcode `\#12\catcode `\^12\catcode `\_12\catcode `\%12\relax}%
\providecommand \@@startlink[1]{}%
\providecommand \@@endlink[0]{}%
\providecommand \url  [0]{\begingroup\@sanitize@url \@url }%
\providecommand \@url [1]{\endgroup\@href {#1}{\urlprefix }}%
\providecommand \urlprefix  [0]{URL }%
\providecommand \Eprint [0]{\href }%
\providecommand \doibase [0]{https://doi.org/}%
\providecommand \selectlanguage [0]{\@gobble}%
\providecommand \bibinfo  [0]{\@secondoftwo}%
\providecommand \bibfield  [0]{\@secondoftwo}%
\providecommand \translation [1]{[#1]}%
\providecommand \BibitemOpen [0]{}%
\providecommand \bibitemStop [0]{}%
\providecommand \bibitemNoStop [0]{.\EOS\space}%
\providecommand \EOS [0]{\spacefactor3000\relax}%
\providecommand \BibitemShut  [1]{\csname bibitem#1\endcsname}%
\let\auto@bib@innerbib\@empty
\bibitem [{\citenamefont {Evans}\ and\ \citenamefont {Xie}(2008)}]{Evans2008}%
  \BibitemOpen
  \bibfield  {author} {\bibinfo {author} {\bibfnamefont {C.~L.}\ \bibnamefont
  {Evans}}\ and\ \bibinfo {author} {\bibfnamefont {X.~S.}\ \bibnamefont
  {Xie}},\ }\bibfield  {title} {\bibinfo {title} {Coherent anti-{S}tokes
  {R}aman {S}cattering {M}icroscopy: Chemical imaging for biology and
  medicine},\ }\href
  {https://doi.org/https://doi.org/10.1146/annurev.anchem.1.031207.112754}
  {\bibfield  {journal} {\bibinfo  {journal} {Annual Review of Analytical
  Chemistry}\ }\textbf {\bibinfo {volume} {1}},\ \bibinfo {pages} {883}
  (\bibinfo {year} {2008})}\BibitemShut {NoStop}%
\bibitem [{\citenamefont {Cheng}\ and\ \citenamefont {Xie}(2015)}]{Cheng2015}%
  \BibitemOpen
  \bibfield  {author} {\bibinfo {author} {\bibfnamefont {J.-X.}\ \bibnamefont
  {Cheng}}\ and\ \bibinfo {author} {\bibfnamefont {X.~S.}\ \bibnamefont
  {Xie}},\ }\bibfield  {title} {\bibinfo {title} {Vibrational spectroscopic
  imaging of living systems: An emerging platform for biology and medicine},\
  }\href {https://doi.org/10.1126/science.aaa8870} {\bibfield  {journal}
  {\bibinfo  {journal} {Science}\ }\textbf {\bibinfo {volume} {350}},\ \bibinfo
  {pages} {aaa8870} (\bibinfo {year} {2015})}\BibitemShut {NoStop}%
\bibitem [{\citenamefont {Voronine}\ \emph {et~al.}(2018)\citenamefont
  {Voronine}, \citenamefont {Zhang}, \citenamefont {Sokolov},\ and\
  \citenamefont {Scully}}]{VoronineZhangSokolovScully+2018+523+548}%
  \BibitemOpen
  \bibfield  {author} {\bibinfo {author} {\bibfnamefont {D.~V.}\ \bibnamefont
  {Voronine}}, \bibinfo {author} {\bibfnamefont {Z.}~\bibnamefont {Zhang}},
  \bibinfo {author} {\bibfnamefont {A.~V.}\ \bibnamefont {Sokolov}},\ and\
  \bibinfo {author} {\bibfnamefont {M.~O.}\ \bibnamefont {Scully}},\ }\bibfield
   {title} {\bibinfo {title} {Surface-enhanced {FAST} {CARS}: en route to
  quantum nano-biophotonics},\ }\href
  {https://doi.org/doi:10.1515/nanoph-2017-0066} {\bibfield  {journal}
  {\bibinfo  {journal} {Nanophotonics}\ }\textbf {\bibinfo {volume} {7}},\
  \bibinfo {pages} {523} (\bibinfo {year} {2018})}\BibitemShut {NoStop}%
\bibitem [{\citenamefont {Camp~Jr}\ and\ \citenamefont
  {Cicerone}(2015)}]{CampJr2015}%
  \BibitemOpen
  \bibfield  {author} {\bibinfo {author} {\bibfnamefont {C.~H.}\ \bibnamefont
  {Camp~Jr}}\ and\ \bibinfo {author} {\bibfnamefont {M.~T.}\ \bibnamefont
  {Cicerone}},\ }\bibfield  {title} {\bibinfo {title} {Chemically sensitive
  bioimaging with coherent {R}aman scattering},\ }\href
  {https://doi.org/10.1038/nphoton.2015.60} {\bibfield  {journal} {\bibinfo
  {journal} {Nature Photonics}\ }\textbf {\bibinfo {volume} {9}},\ \bibinfo
  {pages} {295} (\bibinfo {year} {2015})}\BibitemShut {NoStop}%
\bibitem [{\citenamefont {Prince}\ \emph {et~al.}(2017)\citenamefont {Prince},
  \citenamefont {Frontiera},\ and\ \citenamefont {Potma}}]{Prince2017}%
  \BibitemOpen
  \bibfield  {author} {\bibinfo {author} {\bibfnamefont {R.~C.}\ \bibnamefont
  {Prince}}, \bibinfo {author} {\bibfnamefont {R.~R.}\ \bibnamefont
  {Frontiera}},\ and\ \bibinfo {author} {\bibfnamefont {E.~O.}\ \bibnamefont
  {Potma}},\ }\bibfield  {title} {\bibinfo {title} {Stimulated {R}aman
  {S}cattering: From bulk to nano},\ }\href
  {https://doi.org/10.1021/acs.chemrev.6b00545} {\bibfield  {journal} {\bibinfo
   {journal} {Chemical Reviews}\ }\textbf {\bibinfo {volume} {117}},\ \bibinfo
  {pages} {5070} (\bibinfo {year} {2017})}\BibitemShut {NoStop}%
\bibitem [{\citenamefont {Tipping}\ \emph {et~al.}(2024)\citenamefont
  {Tipping}, \citenamefont {Faulds},\ and\ \citenamefont
  {Graham}}]{Tipping2024}%
  \BibitemOpen
  \bibfield  {author} {\bibinfo {author} {\bibfnamefont {W.~J.}\ \bibnamefont
  {Tipping}}, \bibinfo {author} {\bibfnamefont {K.}~\bibnamefont {Faulds}},\
  and\ \bibinfo {author} {\bibfnamefont {D.}~\bibnamefont {Graham}},\
  }\bibfield  {title} {\bibinfo {title} {Advances in super-resolution
  stimulated {Raman} scattering microscopy},\ }\href
  {https://doi.org/10.1021/cbmi.4c00057} {\bibfield  {journal} {\bibinfo
  {journal} {Chemical {\&} Biomedical Imaging}\ }\textbf {\bibinfo {volume}
  {2}},\ \bibinfo {pages} {733} (\bibinfo {year} {2024})}\BibitemShut {NoStop}%
\bibitem [{\citenamefont {Kukura}\ \emph {et~al.}(2007)\citenamefont {Kukura},
  \citenamefont {McCamant},\ and\ \citenamefont {Mathies}}]{Kukura2007}%
  \BibitemOpen
  \bibfield  {author} {\bibinfo {author} {\bibfnamefont {P.}~\bibnamefont
  {Kukura}}, \bibinfo {author} {\bibfnamefont {D.~W.}\ \bibnamefont
  {McCamant}},\ and\ \bibinfo {author} {\bibfnamefont {R.~A.}\ \bibnamefont
  {Mathies}},\ }\bibfield  {title} {\bibinfo {title} {Femtosecond stimulated
  {{Raman}} spectroscopy},\ }\href
  {https://doi.org/https://doi.org/10.1146/annurev.physchem.58.032806.104456}
  {\bibfield  {journal} {\bibinfo  {journal} {Annual Review of Physical
  Chemistry}\ }\textbf {\bibinfo {volume} {58}},\ \bibinfo {pages} {461}
  (\bibinfo {year} {2007})}\BibitemShut {NoStop}%
\bibitem [{\citenamefont {Batignani}\ \emph {et~al.}(2024)\citenamefont
  {Batignani}, \citenamefont {Ferrante}, \citenamefont {Fumero}, \citenamefont
  {Martinati},\ and\ \citenamefont {Scopigno}}]{Batignani2024}%
  \BibitemOpen
  \bibfield  {author} {\bibinfo {author} {\bibfnamefont {G.}~\bibnamefont
  {Batignani}}, \bibinfo {author} {\bibfnamefont {C.}~\bibnamefont {Ferrante}},
  \bibinfo {author} {\bibfnamefont {G.}~\bibnamefont {Fumero}}, \bibinfo
  {author} {\bibfnamefont {M.}~\bibnamefont {Martinati}},\ and\ \bibinfo
  {author} {\bibfnamefont {T.}~\bibnamefont {Scopigno}},\ }\bibfield  {title}
  {\bibinfo {title} {Femtosecond stimulated {{Raman}} spectroscopy},\ }\href
  {https://doi.org/10.1038/s43586-024-00314-6} {\bibfield  {journal} {\bibinfo
  {journal} {Nature Reviews Methods Primers}\ }\textbf {\bibinfo {volume}
  {4}},\ \bibinfo {pages} {34} (\bibinfo {year} {2024})}\BibitemShut {NoStop}%
\bibitem [{\citenamefont {Pellatz}\ \emph {et~al.}(2021)\citenamefont
  {Pellatz}, \citenamefont {Roy}, \citenamefont {Lee}, \citenamefont {Schad},
  \citenamefont {Kandel}, \citenamefont {Arndt}, \citenamefont {Eom},
  \citenamefont {Kemper},\ and\ \citenamefont {Reznik}}]{Pellatz2021}%
  \BibitemOpen
  \bibfield  {author} {\bibinfo {author} {\bibfnamefont {N.}~\bibnamefont
  {Pellatz}}, \bibinfo {author} {\bibfnamefont {S.}~\bibnamefont {Roy}},
  \bibinfo {author} {\bibfnamefont {J.-W.}\ \bibnamefont {Lee}}, \bibinfo
  {author} {\bibfnamefont {J.~L.}\ \bibnamefont {Schad}}, \bibinfo {author}
  {\bibfnamefont {H.}~\bibnamefont {Kandel}}, \bibinfo {author} {\bibfnamefont
  {N.}~\bibnamefont {Arndt}}, \bibinfo {author} {\bibfnamefont {C.~B.}\
  \bibnamefont {Eom}}, \bibinfo {author} {\bibfnamefont {A.~F.}\ \bibnamefont
  {Kemper}},\ and\ \bibinfo {author} {\bibfnamefont {D.}~\bibnamefont
  {Reznik}},\ }\bibfield  {title} {\bibinfo {title} {Relaxation timescales and
  electron-phonon coupling in optically pumped {YBa}$_{2}${Cu}$_{3}${O}$_{6+x}$
  revealed by time-resolved {Raman} scattering},\ }\href
  {https://doi.org/10.1103/PhysRevB.104.L180505} {\bibfield  {journal}
  {\bibinfo  {journal} {Phys. Rev. B}\ }\textbf {\bibinfo {volume} {104}},\
  \bibinfo {pages} {L180505} (\bibinfo {year} {2021})}\BibitemShut {NoStop}%
\bibitem [{\citenamefont {Chou}\ \emph {et~al.}(2024)\citenamefont {Chou},
  \citenamefont {F\"orst}, \citenamefont {Fechner}, \citenamefont {Henstridge},
  \citenamefont {Roy}, \citenamefont {Buzzi}, \citenamefont {Nicoletti},
  \citenamefont {Liu}, \citenamefont {Nakata}, \citenamefont {Keimer},\ and\
  \citenamefont {Cavalleri}}]{Chou2024}%
  \BibitemOpen
  \bibfield  {author} {\bibinfo {author} {\bibfnamefont {T.-H.}\ \bibnamefont
  {Chou}}, \bibinfo {author} {\bibfnamefont {M.}~\bibnamefont {F\"orst}},
  \bibinfo {author} {\bibfnamefont {M.}~\bibnamefont {Fechner}}, \bibinfo
  {author} {\bibfnamefont {M.}~\bibnamefont {Henstridge}}, \bibinfo {author}
  {\bibfnamefont {S.}~\bibnamefont {Roy}}, \bibinfo {author} {\bibfnamefont
  {M.}~\bibnamefont {Buzzi}}, \bibinfo {author} {\bibfnamefont
  {D.}~\bibnamefont {Nicoletti}}, \bibinfo {author} {\bibfnamefont
  {Y.}~\bibnamefont {Liu}}, \bibinfo {author} {\bibfnamefont {S.}~\bibnamefont
  {Nakata}}, \bibinfo {author} {\bibfnamefont {B.}~\bibnamefont {Keimer}},\
  and\ \bibinfo {author} {\bibfnamefont {A.}~\bibnamefont {Cavalleri}},\
  }\bibfield  {title} {\bibinfo {title} {Ultrafast {Raman} thermometry in
  driven {YBa}$_{2}${Cu}$_{3}${O}$_{6.48}$},\ }\href
  {https://doi.org/10.1103/PhysRevB.109.195141} {\bibfield  {journal} {\bibinfo
   {journal} {Phys. Rev. B}\ }\textbf {\bibinfo {volume} {109}},\ \bibinfo
  {pages} {195141} (\bibinfo {year} {2024})}\BibitemShut {NoStop}%
\bibitem [{\citenamefont {Rahav}\ and\ \citenamefont
  {Mukamel}(2010)}]{Rahav2010}%
  \BibitemOpen
  \bibfield  {author} {\bibinfo {author} {\bibfnamefont {S.}~\bibnamefont
  {Rahav}}\ and\ \bibinfo {author} {\bibfnamefont {S.}~\bibnamefont
  {Mukamel}},\ }\bibfield  {title} {\bibinfo {title} {Stimulated coherent
  anti-stokes {Raman} spectroscopy (cars) resonances originate from double-slit
  interference of two-photon stokes pathways},\ }\href
  {https://doi.org/10.1073/pnas.0910120107} {\bibfield  {journal} {\bibinfo
  {journal} {Proceedings of the National Academy of Sciences}\ }\textbf
  {\bibinfo {volume} {107}},\ \bibinfo {pages} {4825} (\bibinfo {year}
  {2010})}\BibitemShut {NoStop}%
\bibitem [{\citenamefont {Potma}\ and\ \citenamefont
  {Mukamel}(2012)}]{Raman-book}%
  \BibitemOpen
  \bibfield  {author} {\bibinfo {author} {\bibfnamefont {E.~O.}\ \bibnamefont
  {Potma}}\ and\ \bibinfo {author} {\bibfnamefont {S.}~\bibnamefont
  {Mukamel}},\ }\href {https://doi.org/10.1201/b12907} {\emph {\bibinfo {title}
  {Theory of Coherent {{Raman}} Scattering. in: Coherent {{Raman}} Scattering
  Microscopy}}},\ edited by\ \bibinfo {editor} {\bibfnamefont {J.-X.}\
  \bibnamefont {Cheng}}\ and\ \bibinfo {editor} {\bibfnamefont
  {X.}~\bibnamefont {Xie}}\ (\bibinfo  {publisher} {CRC Press},\ \bibinfo
  {year} {2012})\BibitemShut {NoStop}%
\bibitem [{\citenamefont {Bowen}\ \emph {et~al.}(2023)\citenamefont {Bowen},
  \citenamefont {Chrzanowski}, \citenamefont {Oron}, \citenamefont {Ramelow},
  \citenamefont {Tabakaev}, \citenamefont {Terrasson},\ and\ \citenamefont
  {Thew}}]{Bowen-review}%
  \BibitemOpen
  \bibfield  {author} {\bibinfo {author} {\bibfnamefont {W.~P.}\ \bibnamefont
  {Bowen}}, \bibinfo {author} {\bibfnamefont {H.~M.}\ \bibnamefont
  {Chrzanowski}}, \bibinfo {author} {\bibfnamefont {D.}~\bibnamefont {Oron}},
  \bibinfo {author} {\bibfnamefont {S.}~\bibnamefont {Ramelow}}, \bibinfo
  {author} {\bibfnamefont {D.}~\bibnamefont {Tabakaev}}, \bibinfo {author}
  {\bibfnamefont {A.}~\bibnamefont {Terrasson}},\ and\ \bibinfo {author}
  {\bibfnamefont {R.}~\bibnamefont {Thew}},\ }\bibfield  {title} {\bibinfo
  {title} {Quantum light microscopy},\ }\href
  {https://doi.org/10.1080/00107514.2023.2292380} {\bibfield  {journal}
  {\bibinfo  {journal} {Contemporary Physics}\ }\textbf {\bibinfo {volume}
  {64}},\ \bibinfo {pages} {169} (\bibinfo {year} {2023})}\BibitemShut
  {NoStop}%
\bibitem [{\citenamefont {Giovannetti}\ \emph {et~al.}(2011)\citenamefont
  {Giovannetti}, \citenamefont {Lloyd},\ and\ \citenamefont
  {Maccone}}]{VittorioNP2011}%
  \BibitemOpen
  \bibfield  {author} {\bibinfo {author} {\bibfnamefont {V.}~\bibnamefont
  {Giovannetti}}, \bibinfo {author} {\bibfnamefont {S.}~\bibnamefont {Lloyd}},\
  and\ \bibinfo {author} {\bibfnamefont {L.}~\bibnamefont {Maccone}},\
  }\bibfield  {title} {\bibinfo {title} {Advances in quantum metrology},\
  }\href {https://doi.org/10.1038/nphoton.2011.35} {\bibfield  {journal}
  {\bibinfo  {journal} {Nat. Photon.}\ }\textbf {\bibinfo {volume} {5}},\
  \bibinfo {pages} {222} (\bibinfo {year} {2011})}\BibitemShut {NoStop}%
\bibitem [{\citenamefont {Pezz{\`e}}\ and\ \citenamefont
  {Smerzi}(2014)}]{Pezze2014}%
  \BibitemOpen
  \bibfield  {author} {\bibinfo {author} {\bibfnamefont {L.}~\bibnamefont
  {Pezz{\`e}}}\ and\ \bibinfo {author} {\bibfnamefont {A.}~\bibnamefont
  {Smerzi}},\ }\bibfield  {title} {\bibinfo {title} {Quantum theory of phase
  estimation},\ }in\ \href@noop {} {\emph {\bibinfo {booktitle} {Atom
  Interferometry}}}\ (\bibinfo  {publisher} {IOS Press},\ \bibinfo {year}
  {2014})\ pp.\ \bibinfo {pages} {691--741}\BibitemShut {NoStop}%
\bibitem [{\citenamefont {Tsang}\ \emph {et~al.}(2016)\citenamefont {Tsang},
  \citenamefont {Nair},\ and\ \citenamefont {Lu}}]{PhysRevX.6.031033}%
  \BibitemOpen
  \bibfield  {author} {\bibinfo {author} {\bibfnamefont {M.}~\bibnamefont
  {Tsang}}, \bibinfo {author} {\bibfnamefont {R.}~\bibnamefont {Nair}},\ and\
  \bibinfo {author} {\bibfnamefont {X.-M.}\ \bibnamefont {Lu}},\ }\bibfield
  {title} {\bibinfo {title} {Quantum theory of superresolution for two
  incoherent optical point sources},\ }\href
  {https://doi.org/10.1103/PhysRevX.6.031033} {\bibfield  {journal} {\bibinfo
  {journal} {Phys. Rev. X}\ }\textbf {\bibinfo {volume} {6}},\ \bibinfo {pages}
  {031033} (\bibinfo {year} {2016})}\BibitemShut {NoStop}%
\bibitem [{\citenamefont {Tsang}(2019)}]{TsangReview}%
  \BibitemOpen
  \bibfield  {author} {\bibinfo {author} {\bibfnamefont {M.}~\bibnamefont
  {Tsang}},\ }\bibfield  {title} {\bibinfo {title} {Resolving starlight: a
  quantum perspective},\ }\href {https://doi.org/10.1080/00107514.2020.1736375}
  {\bibfield  {journal} {\bibinfo  {journal} {Contemporary Physics}\ }\textbf
  {\bibinfo {volume} {60}},\ \bibinfo {pages} {279} (\bibinfo {year}
  {2019})}\BibitemShut {NoStop}%
\bibitem [{\citenamefont {Caves}(1981)}]{Caves1981}%
  \BibitemOpen
  \bibfield  {author} {\bibinfo {author} {\bibfnamefont {C.~M.}\ \bibnamefont
  {Caves}},\ }\bibfield  {title} {\bibinfo {title} {Quantum-mechanical noise in
  an interferometer},\ }\href {https://doi.org/10.1103/PhysRevD.23.1693}
  {\bibfield  {journal} {\bibinfo  {journal} {Phys. Rev. D}\ }\textbf {\bibinfo
  {volume} {23}},\ \bibinfo {pages} {1693} (\bibinfo {year}
  {1981})}\BibitemShut {NoStop}%
\bibitem [{\citenamefont {Collaboration}(2019)}]{VirgoPRL2019}%
  \BibitemOpen
  \bibfield  {author} {\bibinfo {author} {\bibfnamefont {V.}~\bibnamefont
  {Collaboration}},\ }\bibfield  {title} {\bibinfo {title} {Increasing the
  astrophysical reach of the advanced virgo detector via the application of
  squeezed vacuum states of light},\ }\href
  {https://doi.org/10.1103/PhysRevLett.123.231108} {\bibfield  {journal}
  {\bibinfo  {journal} {Phys. Rev. Lett.}\ }\textbf {\bibinfo {volume} {123}},\
  \bibinfo {pages} {231108} (\bibinfo {year} {2019})}\BibitemShut {NoStop}%
\bibitem [{\citenamefont {Tse}\ \emph {et~al.}(2019)\citenamefont {Tse},
  \citenamefont {Yu}, \citenamefont {Kijbunchoo},\ and\ \citenamefont
  {et~al.}}]{TsePRL2019}%
  \BibitemOpen
  \bibfield  {author} {\bibinfo {author} {\bibfnamefont {M.}~\bibnamefont
  {Tse}}, \bibinfo {author} {\bibfnamefont {H.}~\bibnamefont {Yu}}, \bibinfo
  {author} {\bibfnamefont {N.}~\bibnamefont {Kijbunchoo}},\ and\ \bibinfo
  {author} {\bibnamefont {et~al.}},\ }\bibfield  {title} {\bibinfo {title}
  {Quantum-enhanced advanced {LIGO} detectors in the era of gravitational-wave
  astronomy},\ }\href {https://doi.org/10.1103/PhysRevLett.123.231107}
  {\bibfield  {journal} {\bibinfo  {journal} {Phys. Rev. Lett.}\ }\textbf
  {\bibinfo {volume} {123}},\ \bibinfo {pages} {231107} (\bibinfo {year}
  {2019})}\BibitemShut {NoStop}%
\bibitem [{\citenamefont {Dorfman}\ \emph {et~al.}(2014)\citenamefont
  {Dorfman}, \citenamefont {Schlawin},\ and\ \citenamefont
  {Mukamel}}]{Dorfman2014}%
  \BibitemOpen
  \bibfield  {author} {\bibinfo {author} {\bibfnamefont {K.~E.}\ \bibnamefont
  {Dorfman}}, \bibinfo {author} {\bibfnamefont {F.}~\bibnamefont {Schlawin}},\
  and\ \bibinfo {author} {\bibfnamefont {S.}~\bibnamefont {Mukamel}},\
  }\bibfield  {title} {\bibinfo {title} {Stimulated {Raman} spectroscopy with
  entangled light: Enhanced resolution and pathway selection},\ }\href
  {https://doi.org/10.1021/jz501124a} {\bibfield  {journal} {\bibinfo
  {journal} {The Journal of Physical Chemistry Letters}\ }\textbf {\bibinfo
  {volume} {5}},\ \bibinfo {pages} {2843} (\bibinfo {year} {2014})}\BibitemShut
  {NoStop}%
\bibitem [{\citenamefont {Munkhbaatar}\ and\ \citenamefont
  {Myung-Whun}(2017)}]{MUNKHBAATAR2017581}%
  \BibitemOpen
  \bibfield  {author} {\bibinfo {author} {\bibfnamefont {P.}~\bibnamefont
  {Munkhbaatar}}\ and\ \bibinfo {author} {\bibfnamefont {K.}~\bibnamefont
  {Myung-Whun}},\ }\bibfield  {title} {\bibinfo {title} {Selection of
  stimulated {Raman} scattering signal by entangled photons},\ }\href
  {https://doi.org/https://doi.org/10.1016/j.optcom.2016.09.061} {\bibfield
  {journal} {\bibinfo  {journal} {Optics Communications}\ }\textbf {\bibinfo
  {volume} {383}},\ \bibinfo {pages} {581} (\bibinfo {year}
  {2017})}\BibitemShut {NoStop}%
\bibitem [{\citenamefont {Chen}\ and\ \citenamefont
  {Mukamel}(2021)}]{Chen2021}%
  \BibitemOpen
  \bibfield  {author} {\bibinfo {author} {\bibfnamefont {F.}~\bibnamefont
  {Chen}}\ and\ \bibinfo {author} {\bibfnamefont {S.}~\bibnamefont {Mukamel}},\
  }\bibfield  {title} {\bibinfo {title} {Vibrational hyper-{Raman} molecular
  spectroscopy with entangled photons},\ }\href
  {https://doi.org/10.1021/acsphotonics.1c00777} {\bibfield  {journal}
  {\bibinfo  {journal} {ACS Photonics}\ }\textbf {\bibinfo {volume} {8}},\
  \bibinfo {pages} {2722} (\bibinfo {year} {2021})}\BibitemShut {NoStop}%
\bibitem [{\citenamefont {Schlawin}\ \emph {et~al.}(2021)\citenamefont
  {Schlawin}, \citenamefont {Dorfman},\ and\ \citenamefont
  {Mukamel}}]{Schlawin2021}%
  \BibitemOpen
  \bibfield  {author} {\bibinfo {author} {\bibfnamefont {F.}~\bibnamefont
  {Schlawin}}, \bibinfo {author} {\bibfnamefont {K.~E.}\ \bibnamefont
  {Dorfman}},\ and\ \bibinfo {author} {\bibfnamefont {S.}~\bibnamefont
  {Mukamel}},\ }\bibfield  {title} {\bibinfo {title} {{Detection of photon
  statistics and multimode field correlations by {Raman} processes}},\ }\href
  {https://doi.org/10.1063/5.0039759} {\bibfield  {journal} {\bibinfo
  {journal} {The Journal of Chemical Physics}\ }\textbf {\bibinfo {volume}
  {154}},\ \bibinfo {pages} {104116} (\bibinfo {year} {2021})}\BibitemShut
  {NoStop}%
\bibitem [{\citenamefont {Fan}\ \emph {et~al.}(2024)\citenamefont {Fan},
  \citenamefont {Ou},\ and\ \citenamefont {Zhang}}]{Fan2024}%
  \BibitemOpen
  \bibfield  {author} {\bibinfo {author} {\bibfnamefont {J.~J.}\ \bibnamefont
  {Fan}}, \bibinfo {author} {\bibfnamefont {Z.-Y.}\ \bibnamefont {Ou}},\ and\
  \bibinfo {author} {\bibfnamefont {Z.}~\bibnamefont {Zhang}},\ }\bibfield
  {title} {\bibinfo {title} {Entangled photons enabled ultrafast stimulated
  {Raman} spectroscopy for molecular dynamics},\ }\href
  {https://doi.org/10.1038/s41377-024-01492-4} {\bibfield  {journal} {\bibinfo
  {journal} {Light: Science {\&} Applications}\ }\textbf {\bibinfo {volume}
  {13}},\ \bibinfo {pages} {163} (\bibinfo {year} {2024})}\BibitemShut
  {NoStop}%
\bibitem [{\citenamefont {Jadoun}\ \emph {et~al.}(2024)\citenamefont {Jadoun},
  \citenamefont {Zhang},\ and\ \citenamefont {Kowalewski}}]{Jadoun2024}%
  \BibitemOpen
  \bibfield  {author} {\bibinfo {author} {\bibfnamefont {D.}~\bibnamefont
  {Jadoun}}, \bibinfo {author} {\bibfnamefont {Z.}~\bibnamefont {Zhang}},\ and\
  \bibinfo {author} {\bibfnamefont {M.}~\bibnamefont {Kowalewski}},\ }\bibfield
   {title} {\bibinfo {title} {{Raman} spectroscopy of conical intersections
  using entangled photons},\ }\href
  {https://doi.org/10.1021/acs.jpclett.3c02852} {\bibfield  {journal} {\bibinfo
   {journal} {The Journal of Physical Chemistry Letters}\ }\textbf {\bibinfo
  {volume} {15}},\ \bibinfo {pages} {2023} (\bibinfo {year}
  {2024})}\BibitemShut {NoStop}%
\bibitem [{\citenamefont {de~Andrade}\ \emph {et~al.}(2020)\citenamefont
  {de~Andrade}, \citenamefont {Kerdoncuff}, \citenamefont {Berg-S{\o}rensen},
  \citenamefont {Gehring}, \citenamefont {Lassen},\ and\ \citenamefont
  {Andersen}}]{deAndrade:20}%
  \BibitemOpen
  \bibfield  {author} {\bibinfo {author} {\bibfnamefont {R.~B.}\ \bibnamefont
  {de~Andrade}}, \bibinfo {author} {\bibfnamefont {H.}~\bibnamefont
  {Kerdoncuff}}, \bibinfo {author} {\bibfnamefont {K.}~\bibnamefont
  {Berg-S{\o}rensen}}, \bibinfo {author} {\bibfnamefont {T.}~\bibnamefont
  {Gehring}}, \bibinfo {author} {\bibfnamefont {M.}~\bibnamefont {Lassen}},\
  and\ \bibinfo {author} {\bibfnamefont {U.~L.}\ \bibnamefont {Andersen}},\
  }\bibfield  {title} {\bibinfo {title} {Quantum-enhanced continuous-wave
  stimulated {Raman} scattering spectroscopy},\ }\href
  {https://doi.org/10.1364/OPTICA.386584} {\bibfield  {journal} {\bibinfo
  {journal} {Optica}\ }\textbf {\bibinfo {volume} {7}},\ \bibinfo {pages} {470}
  (\bibinfo {year} {2020})}\BibitemShut {NoStop}%
\bibitem [{\citenamefont {Casacio}\ \emph {et~al.}(2021)\citenamefont
  {Casacio}, \citenamefont {Madsen}, \citenamefont {Terrasson}, \citenamefont
  {Waleed}, \citenamefont {Barnscheidt}, \citenamefont {Hage}, \citenamefont
  {Taylor},\ and\ \citenamefont {Bowen}}]{Casacio2021}%
  \BibitemOpen
  \bibfield  {author} {\bibinfo {author} {\bibfnamefont {C.~A.}\ \bibnamefont
  {Casacio}}, \bibinfo {author} {\bibfnamefont {L.~S.}\ \bibnamefont {Madsen}},
  \bibinfo {author} {\bibfnamefont {A.}~\bibnamefont {Terrasson}}, \bibinfo
  {author} {\bibfnamefont {M.}~\bibnamefont {Waleed}}, \bibinfo {author}
  {\bibfnamefont {K.}~\bibnamefont {Barnscheidt}}, \bibinfo {author}
  {\bibfnamefont {B.}~\bibnamefont {Hage}}, \bibinfo {author} {\bibfnamefont
  {M.~A.}\ \bibnamefont {Taylor}},\ and\ \bibinfo {author} {\bibfnamefont
  {W.~P.}\ \bibnamefont {Bowen}},\ }\bibfield  {title} {\bibinfo {title}
  {Quantum-enhanced nonlinear microscopy},\ }\href
  {https://doi.org/10.1038/s41586-021-03528-w} {\bibfield  {journal} {\bibinfo
  {journal} {Nature}\ }\textbf {\bibinfo {volume} {594}},\ \bibinfo {pages}
  {201} (\bibinfo {year} {2021})}\BibitemShut {NoStop}%
\bibitem [{\citenamefont {Xu}\ \emph {et~al.}(2022)\citenamefont {Xu},
  \citenamefont {Oguchi}, \citenamefont {Taguchi}, \citenamefont {Takahashi},
  \citenamefont {Sano}, \citenamefont {Mizuguchi}, \citenamefont {Katoh},\ and\
  \citenamefont {Ozeki}}]{Xu22}%
  \BibitemOpen
  \bibfield  {author} {\bibinfo {author} {\bibfnamefont {Z.}~\bibnamefont
  {Xu}}, \bibinfo {author} {\bibfnamefont {K.}~\bibnamefont {Oguchi}}, \bibinfo
  {author} {\bibfnamefont {Y.}~\bibnamefont {Taguchi}}, \bibinfo {author}
  {\bibfnamefont {S.}~\bibnamefont {Takahashi}}, \bibinfo {author}
  {\bibfnamefont {Y.}~\bibnamefont {Sano}}, \bibinfo {author} {\bibfnamefont
  {T.}~\bibnamefont {Mizuguchi}}, \bibinfo {author} {\bibfnamefont
  {K.}~\bibnamefont {Katoh}},\ and\ \bibinfo {author} {\bibfnamefont
  {Y.}~\bibnamefont {Ozeki}},\ }\bibfield  {title} {\bibinfo {title}
  {Quantum-enhanced stimulated {Raman} scattering microscopy in a high-power
  regime},\ }\href {https://doi.org/10.1364/OL.473130} {\bibfield  {journal}
  {\bibinfo  {journal} {Opt. Lett.}\ }\textbf {\bibinfo {volume} {47}},\
  \bibinfo {pages} {5829} (\bibinfo {year} {2022})}\BibitemShut {NoStop}%
\bibitem [{\citenamefont {Terrasson}\ \emph {et~al.}(2024)\citenamefont
  {Terrasson}, \citenamefont {Mauranyapin}, \citenamefont {Casacio},
  \citenamefont {Grim}, \citenamefont {Barnscheidt}, \citenamefont {Hage},
  \citenamefont {Taylor},\ and\ \citenamefont {Bowen}}]{Terrasson24}%
  \BibitemOpen
  \bibfield  {author} {\bibinfo {author} {\bibfnamefont {A.}~\bibnamefont
  {Terrasson}}, \bibinfo {author} {\bibfnamefont {N.~P.}\ \bibnamefont
  {Mauranyapin}}, \bibinfo {author} {\bibfnamefont {C.~A.}\ \bibnamefont
  {Casacio}}, \bibinfo {author} {\bibfnamefont {J.~Q.}\ \bibnamefont {Grim}},
  \bibinfo {author} {\bibfnamefont {K.}~\bibnamefont {Barnscheidt}}, \bibinfo
  {author} {\bibfnamefont {B.}~\bibnamefont {Hage}}, \bibinfo {author}
  {\bibfnamefont {M.~A.}\ \bibnamefont {Taylor}},\ and\ \bibinfo {author}
  {\bibfnamefont {W.~P.}\ \bibnamefont {Bowen}},\ }\bibfield  {title} {\bibinfo
  {title} {Fast biological imaging with quantum-enhanced {Raman} microscopy},\
  }\href {https://doi.org/10.1364/OE.523956} {\bibfield  {journal} {\bibinfo
  {journal} {Opt. Express}\ }\textbf {\bibinfo {volume} {32}},\ \bibinfo
  {pages} {36193} (\bibinfo {year} {2024})}\BibitemShut {NoStop}%
\bibitem [{\citenamefont {Li}\ \emph {et~al.}(2022)\citenamefont {Li},
  \citenamefont {Li}, \citenamefont {Liu}, \citenamefont {Yakovlev},\ and\
  \citenamefont {Agarwal}}]{Li2022}%
  \BibitemOpen
  \bibfield  {author} {\bibinfo {author} {\bibfnamefont {T.}~\bibnamefont
  {Li}}, \bibinfo {author} {\bibfnamefont {F.}~\bibnamefont {Li}}, \bibinfo
  {author} {\bibfnamefont {X.}~\bibnamefont {Liu}}, \bibinfo {author}
  {\bibfnamefont {V.~V.}\ \bibnamefont {Yakovlev}},\ and\ \bibinfo {author}
  {\bibfnamefont {G.~S.}\ \bibnamefont {Agarwal}},\ }\bibfield  {title}
  {\bibinfo {title} {Quantum-enhanced stimulated {Brillouin} scattering
  spectroscopy and imaging},\ }\href {https://doi.org/10.1364/OPTICA.467635}
  {\bibfield  {journal} {\bibinfo  {journal} {Optica}\ }\textbf {\bibinfo
  {volume} {9}},\ \bibinfo {pages} {959} (\bibinfo {year} {2022})}\BibitemShut
  {NoStop}%
\bibitem [{\citenamefont {Li}\ \emph {et~al.}(2024{\natexlab{a}})\citenamefont
  {Li}, \citenamefont {Cheburkanov}, \citenamefont {Yakovlev}, \citenamefont
  {Agarwal},\ and\ \citenamefont {Scully}}]{Li2024}%
  \BibitemOpen
  \bibfield  {author} {\bibinfo {author} {\bibfnamefont {T.}~\bibnamefont
  {Li}}, \bibinfo {author} {\bibfnamefont {V.}~\bibnamefont {Cheburkanov}},
  \bibinfo {author} {\bibfnamefont {V.~V.}\ \bibnamefont {Yakovlev}}, \bibinfo
  {author} {\bibfnamefont {G.~S.}\ \bibnamefont {Agarwal}},\ and\ \bibinfo
  {author} {\bibfnamefont {M.~O.}\ \bibnamefont {Scully}},\ }\bibfield  {title}
  {\bibinfo {title} {Harnessing quantum light for microscopic biomechanical
  imaging of cells and tissues},\ }\href
  {https://doi.org/10.1073/pnas.2413938121} {\bibfield  {journal} {\bibinfo
  {journal} {Proceedings of the National Academy of Sciences}\ }\textbf
  {\bibinfo {volume} {121}},\ \bibinfo {pages} {e2413938121} (\bibinfo {year}
  {2024}{\natexlab{a}})}\BibitemShut {NoStop}%
\bibitem [{\citenamefont {Triginer~Garces}\ \emph {et~al.}(2020)\citenamefont
  {Triginer~Garces}, \citenamefont {Chrzanowski}, \citenamefont {Daryanoosh},
  \citenamefont {Thiel}, \citenamefont {Marchant}, \citenamefont {Patel},
  \citenamefont {Humphreys}, \citenamefont {Datta},\ and\ \citenamefont
  {Walmsley}}]{Triginer-Garces2020}%
  \BibitemOpen
  \bibfield  {author} {\bibinfo {author} {\bibfnamefont {G.}~\bibnamefont
  {Triginer~Garces}}, \bibinfo {author} {\bibfnamefont {H.~M.}\ \bibnamefont
  {Chrzanowski}}, \bibinfo {author} {\bibfnamefont {S.}~\bibnamefont
  {Daryanoosh}}, \bibinfo {author} {\bibfnamefont {V.}~\bibnamefont {Thiel}},
  \bibinfo {author} {\bibfnamefont {A.~L.}\ \bibnamefont {Marchant}}, \bibinfo
  {author} {\bibfnamefont {R.~B.}\ \bibnamefont {Patel}}, \bibinfo {author}
  {\bibfnamefont {P.~C.}\ \bibnamefont {Humphreys}}, \bibinfo {author}
  {\bibfnamefont {A.}~\bibnamefont {Datta}},\ and\ \bibinfo {author}
  {\bibfnamefont {I.~A.}\ \bibnamefont {Walmsley}},\ }\bibfield  {title}
  {\bibinfo {title} {{Quantum-enhanced stimulated emission detection for
  label-free microscopy}},\ }\href {https://doi.org/10.1063/5.0009681}
  {\bibfield  {journal} {\bibinfo  {journal} {Applied Physics Letters}\
  }\textbf {\bibinfo {volume} {117}},\ \bibinfo {pages} {024002} (\bibinfo
  {year} {2020})}\BibitemShut {NoStop}%
\bibitem [{\citenamefont {Gong}\ \emph {et~al.}(2023)\citenamefont {Gong},
  \citenamefont {Lin},\ and\ \citenamefont {Huang}}]{Gong23}%
  \BibitemOpen
  \bibfield  {author} {\bibinfo {author} {\bibfnamefont {L.}~\bibnamefont
  {Gong}}, \bibinfo {author} {\bibfnamefont {S.}~\bibnamefont {Lin}},\ and\
  \bibinfo {author} {\bibfnamefont {Z.}~\bibnamefont {Huang}},\ }\bibfield
  {title} {\bibinfo {title} {Super-resolution stimulated {Raman} scattering
  microscopy enhanced by quantum light and deconvolution},\ }\href
  {https://doi.org/10.1364/OL.509616} {\bibfield  {journal} {\bibinfo
  {journal} {Opt. Lett.}\ }\textbf {\bibinfo {volume} {48}},\ \bibinfo {pages}
  {6516} (\bibinfo {year} {2023})}\BibitemShut {NoStop}%
\bibitem [{\citenamefont {Michael}\ \emph {et~al.}(2019)\citenamefont
  {Michael}, \citenamefont {Bello}, \citenamefont {Rosenbluh},\ and\
  \citenamefont {Pe'er}}]{Michael2019}%
  \BibitemOpen
  \bibfield  {author} {\bibinfo {author} {\bibfnamefont {Y.}~\bibnamefont
  {Michael}}, \bibinfo {author} {\bibfnamefont {L.}~\bibnamefont {Bello}},
  \bibinfo {author} {\bibfnamefont {M.}~\bibnamefont {Rosenbluh}},\ and\
  \bibinfo {author} {\bibfnamefont {A.}~\bibnamefont {Pe'er}},\ }\bibfield
  {title} {\bibinfo {title} {Squeezing-enhanced {Raman} spectroscopy},\ }\href
  {https://doi.org/10.1038/s41534-019-0197-0} {\bibfield  {journal} {\bibinfo
  {journal} {npj Quantum Information}\ }\textbf {\bibinfo {volume} {5}},\
  \bibinfo {pages} {81} (\bibinfo {year} {2019})}\BibitemShut {NoStop}%
\bibitem [{\citenamefont {Li}\ \emph {et~al.}(2024{\natexlab{b}})\citenamefont
  {Li}, \citenamefont {Kumar}, \citenamefont {Huo}, \citenamefont {Du},\ and\
  \citenamefont {Huang}}]{Li24}%
  \BibitemOpen
  \bibfield  {author} {\bibinfo {author} {\bibfnamefont {Y.}~\bibnamefont
  {Li}}, \bibinfo {author} {\bibfnamefont {S.}~\bibnamefont {Kumar}}, \bibinfo
  {author} {\bibfnamefont {T.}~\bibnamefont {Huo}}, \bibinfo {author}
  {\bibfnamefont {H.}~\bibnamefont {Du}},\ and\ \bibinfo {author}
  {\bibfnamefont {Y.-P.}\ \bibnamefont {Huang}},\ }\bibfield  {title} {\bibinfo
  {title} {Photon counting {Raman} spectroscopy: a benchmarking study vs
  surface plasmon enhancement},\ }\href {https://doi.org/10.1364/OE.516970}
  {\bibfield  {journal} {\bibinfo  {journal} {Opt. Express}\ }\textbf {\bibinfo
  {volume} {32}},\ \bibinfo {pages} {16657} (\bibinfo {year}
  {2024}{\natexlab{b}})}\BibitemShut {NoStop}%
\bibitem [{\citenamefont {Xavier}\ \emph {et~al.}(2021)\citenamefont {Xavier},
  \citenamefont {Yu}, \citenamefont {Jones}, \citenamefont {Zossimova},\ and\
  \citenamefont {Vollmer}}]{XavierYuJonesZossimovaVollmer+2021+1387+1435}%
  \BibitemOpen
  \bibfield  {author} {\bibinfo {author} {\bibfnamefont {J.}~\bibnamefont
  {Xavier}}, \bibinfo {author} {\bibfnamefont {D.}~\bibnamefont {Yu}}, \bibinfo
  {author} {\bibfnamefont {C.}~\bibnamefont {Jones}}, \bibinfo {author}
  {\bibfnamefont {E.}~\bibnamefont {Zossimova}},\ and\ \bibinfo {author}
  {\bibfnamefont {F.}~\bibnamefont {Vollmer}},\ }\bibfield  {title} {\bibinfo
  {title} {Quantum nanophotonic and nanoplasmonic sensing: towards quantum
  optical bioscience laboratories on chip},\ }\href
  {https://doi.org/doi:10.1515/nanoph-2020-0593} {\bibfield  {journal}
  {\bibinfo  {journal} {Nanophotonics}\ }\textbf {\bibinfo {volume} {10}},\
  \bibinfo {pages} {1387} (\bibinfo {year} {2021})}\BibitemShut {NoStop}%
\bibitem [{\citenamefont {Gianani}\ \emph {et~al.}(2021)\citenamefont
  {Gianani}, \citenamefont {Albarelli}, \citenamefont {Verna}, \citenamefont
  {Cimini}, \citenamefont {Demkowicz-Dobrzanski},\ and\ \citenamefont
  {Barbieri}}]{Gianani2021}%
  \BibitemOpen
  \bibfield  {author} {\bibinfo {author} {\bibfnamefont {I.}~\bibnamefont
  {Gianani}}, \bibinfo {author} {\bibfnamefont {F.}~\bibnamefont {Albarelli}},
  \bibinfo {author} {\bibfnamefont {A.}~\bibnamefont {Verna}}, \bibinfo
  {author} {\bibfnamefont {V.}~\bibnamefont {Cimini}}, \bibinfo {author}
  {\bibfnamefont {R.}~\bibnamefont {Demkowicz-Dobrzanski}},\ and\ \bibinfo
  {author} {\bibfnamefont {M.}~\bibnamefont {Barbieri}},\ }\bibfield  {title}
  {\bibinfo {title} {Kramers--kronig relations and precision limits in quantum
  phase estimation},\ }\href {https://doi.org/10.1364/OPTICA.440438} {\bibfield
   {journal} {\bibinfo  {journal} {Optica}\ }\textbf {\bibinfo {volume} {8}},\
  \bibinfo {pages} {1642} (\bibinfo {year} {2021})}\BibitemShut {NoStop}%
\bibitem [{\citenamefont {S\'anchez Mu\~noz}\ \emph {et~al.}(2021)\citenamefont
  {S\'anchez Mu\~noz}, \citenamefont {Frascella},\ and\ \citenamefont
  {Schlawin}}]{SanchezMunoz2021}%
  \BibitemOpen
  \bibfield  {author} {\bibinfo {author} {\bibfnamefont {C.}~\bibnamefont
  {S\'anchez Mu\~noz}}, \bibinfo {author} {\bibfnamefont {G.}~\bibnamefont
  {Frascella}},\ and\ \bibinfo {author} {\bibfnamefont {F.}~\bibnamefont
  {Schlawin}},\ }\bibfield  {title} {\bibinfo {title} {Quantum metrology of
  two-photon absorption},\ }\href
  {https://doi.org/10.1103/PhysRevResearch.3.033250} {\bibfield  {journal}
  {\bibinfo  {journal} {Phys. Rev. Res.}\ }\textbf {\bibinfo {volume} {3}},\
  \bibinfo {pages} {033250} (\bibinfo {year} {2021})}\BibitemShut {NoStop}%
\bibitem [{\citenamefont {Panahiyan}\ \emph {et~al.}(2023)\citenamefont
  {Panahiyan}, \citenamefont {Mu\~noz}, \citenamefont {Chekhova},\ and\
  \citenamefont {Schlawin}}]{Panahiyan2023}%
  \BibitemOpen
  \bibfield  {author} {\bibinfo {author} {\bibfnamefont {S.}~\bibnamefont
  {Panahiyan}}, \bibinfo {author} {\bibfnamefont {C.~S.}\ \bibnamefont
  {Mu\~noz}}, \bibinfo {author} {\bibfnamefont {M.~V.}\ \bibnamefont
  {Chekhova}},\ and\ \bibinfo {author} {\bibfnamefont {F.}~\bibnamefont
  {Schlawin}},\ }\bibfield  {title} {\bibinfo {title} {Nonlinear interferometry
  for quantum-enhanced measurements of multiphoton absorption},\ }\href
  {https://doi.org/10.1103/PhysRevLett.130.203604} {\bibfield  {journal}
  {\bibinfo  {journal} {Phys. Rev. Lett.}\ }\textbf {\bibinfo {volume} {130}},\
  \bibinfo {pages} {203604} (\bibinfo {year} {2023})}\BibitemShut {NoStop}%
\bibitem [{\citenamefont {Albarelli}\ \emph {et~al.}(2023)\citenamefont
  {Albarelli}, \citenamefont {Bisketzi}, \citenamefont {Khan},\ and\
  \citenamefont {Datta}}]{Albarelli2023}%
  \BibitemOpen
  \bibfield  {author} {\bibinfo {author} {\bibfnamefont {F.}~\bibnamefont
  {Albarelli}}, \bibinfo {author} {\bibfnamefont {E.}~\bibnamefont {Bisketzi}},
  \bibinfo {author} {\bibfnamefont {A.}~\bibnamefont {Khan}},\ and\ \bibinfo
  {author} {\bibfnamefont {A.}~\bibnamefont {Datta}},\ }\bibfield  {title}
  {\bibinfo {title} {Fundamental limits of pulsed quantum light spectroscopy:
  Dipole moment estimation},\ }\href
  {https://doi.org/10.1103/PhysRevA.107.062601} {\bibfield  {journal} {\bibinfo
   {journal} {Phys. Rev. A}\ }\textbf {\bibinfo {volume} {107}},\ \bibinfo
  {pages} {062601} (\bibinfo {year} {2023})}\BibitemShut {NoStop}%
\bibitem [{\citenamefont {Karsa}\ \emph {et~al.}(2024)\citenamefont {Karsa},
  \citenamefont {Nair}, \citenamefont {Chia}, \citenamefont {Lee},\ and\
  \citenamefont {Lee}}]{Karsa2024}%
  \BibitemOpen
  \bibfield  {author} {\bibinfo {author} {\bibfnamefont {A.}~\bibnamefont
  {Karsa}}, \bibinfo {author} {\bibfnamefont {R.}~\bibnamefont {Nair}},
  \bibinfo {author} {\bibfnamefont {A.}~\bibnamefont {Chia}}, \bibinfo {author}
  {\bibfnamefont {K.-G.}\ \bibnamefont {Lee}},\ and\ \bibinfo {author}
  {\bibfnamefont {C.}~\bibnamefont {Lee}},\ }\bibfield  {title} {\bibinfo
  {title} {Optimal quantum metrology of two-photon absorption},\ }\href
  {https://doi.org/10.1088/2058-9565/ad466b} {\bibfield  {journal} {\bibinfo
  {journal} {Quantum Science and Technology}\ }\textbf {\bibinfo {volume}
  {9}},\ \bibinfo {pages} {035042} (\bibinfo {year} {2024})}\BibitemShut
  {NoStop}%
\bibitem [{\citenamefont {Mukamel}(1999)}]{Shaul_book}%
  \BibitemOpen
  \bibfield  {author} {\bibinfo {author} {\bibfnamefont {S.}~\bibnamefont
  {Mukamel}},\ }\href {https://books.google.co.uk/books?id=garwwAEACAAJ} {\emph
  {\bibinfo {title} {Principles of Nonlinear Optical Spectroscopy}}},\ Oxford
  series in optical and imaging sciences\ (\bibinfo  {publisher} {Oxford
  University Press},\ \bibinfo {address} {Oxford, UK},\ \bibinfo {year}
  {1999})\BibitemShut {NoStop}%
\bibitem [{\citenamefont {Breuer}\ and\ \citenamefont
  {Petruccione}(2007)}]{BreuerPetruccione}%
  \BibitemOpen
  \bibfield  {author} {\bibinfo {author} {\bibfnamefont {H.-P.}\ \bibnamefont
  {Breuer}}\ and\ \bibinfo {author} {\bibfnamefont {F.}~\bibnamefont
  {Petruccione}},\ }\href
  {https://doi.org/10.1093/acprof:oso/9780199213900.001.0001} {\emph {\bibinfo
  {title} {{The Theory of Open Quantum Systems}}}}\ (\bibinfo  {publisher}
  {Oxford University Press},\ \bibinfo {year} {2007})\BibitemShut {NoStop}%
\bibitem [{\citenamefont {Brecht}\ \emph {et~al.}(2015)\citenamefont {Brecht},
  \citenamefont {Reddy}, \citenamefont {Silberhorn},\ and\ \citenamefont
  {Raymer}}]{Brecht2015}%
  \BibitemOpen
  \bibfield  {author} {\bibinfo {author} {\bibfnamefont {B.}~\bibnamefont
  {Brecht}}, \bibinfo {author} {\bibfnamefont {D.~V.}\ \bibnamefont {Reddy}},
  \bibinfo {author} {\bibfnamefont {C.}~\bibnamefont {Silberhorn}},\ and\
  \bibinfo {author} {\bibfnamefont {M.~G.}\ \bibnamefont {Raymer}},\ }\bibfield
   {title} {\bibinfo {title} {Photon temporal modes: A complete framework for
  quantum information science},\ }\href
  {https://doi.org/10.1103/PhysRevX.5.041017} {\bibfield  {journal} {\bibinfo
  {journal} {Phys. Rev. X}\ }\textbf {\bibinfo {volume} {5}},\ \bibinfo {pages}
  {041017} (\bibinfo {year} {2015})}\BibitemShut {NoStop}%
\bibitem [{\citenamefont {Mukamel}(2009)}]{Mukamel2009}%
  \BibitemOpen
  \bibfield  {author} {\bibinfo {author} {\bibfnamefont {S.}~\bibnamefont
  {Mukamel}},\ }\bibfield  {title} {\bibinfo {title} {{Controlling
  multidimensional off-resonant-{Raman} and infrared vibrational spectroscopy
  by finite pulse band shapes}},\ }\href {https://doi.org/10.1063/1.3068548}
  {\bibfield  {journal} {\bibinfo  {journal} {The Journal of Chemical Physics}\
  }\textbf {\bibinfo {volume} {130}},\ \bibinfo {pages} {054110} (\bibinfo
  {year} {2009})}\BibitemShut {NoStop}%
\bibitem [{\citenamefont {Walls}(1970)}]{Walls1970}%
  \BibitemOpen
  \bibfield  {author} {\bibinfo {author} {\bibfnamefont {D.~F.}\ \bibnamefont
  {Walls}},\ }\bibfield  {title} {\bibinfo {title} {Quantum theory of the
  {Raman} effect},\ }\href {https://doi.org/10.1007/BF01398635} {\bibfield
  {journal} {\bibinfo  {journal} {Zeitschrift f{\"u}r Physik A Hadrons and
  nuclei}\ }\textbf {\bibinfo {volume} {237}},\ \bibinfo {pages} {224}
  (\bibinfo {year} {1970})}\BibitemShut {NoStop}%
\bibitem [{\citenamefont {Walls}(1971)}]{Walls1971}%
  \BibitemOpen
  \bibfield  {author} {\bibinfo {author} {\bibfnamefont {D.~F.}\ \bibnamefont
  {Walls}},\ }\bibfield  {title} {\bibinfo {title} {Quantum theory of the
  {Raman} effect},\ }\href {https://doi.org/10.1007/BF01407253} {\bibfield
  {journal} {\bibinfo  {journal} {Zeitschrift f{\"u}r Physik A Hadrons and
  nuclei}\ }\textbf {\bibinfo {volume} {244}},\ \bibinfo {pages} {117}
  (\bibinfo {year} {1971})}\BibitemShut {NoStop}%
\bibitem [{\citenamefont {Raymer}\ and\ \citenamefont
  {Mostowski}(1981)}]{Raymer1980}%
  \BibitemOpen
  \bibfield  {author} {\bibinfo {author} {\bibfnamefont {M.~G.}\ \bibnamefont
  {Raymer}}\ and\ \bibinfo {author} {\bibfnamefont {J.}~\bibnamefont
  {Mostowski}},\ }\bibfield  {title} {\bibinfo {title} {Stimulated {Raman}
  scattering: Unified treatment of spontaneous initiation and spatial
  propagation},\ }\href {https://doi.org/10.1103/PhysRevA.24.1980} {\bibfield
  {journal} {\bibinfo  {journal} {Phys. Rev. A}\ }\textbf {\bibinfo {volume}
  {24}},\ \bibinfo {pages} {1980} (\bibinfo {year} {1981})}\BibitemShut
  {NoStop}%
\bibitem [{\citenamefont {Raymer}\ \emph {et~al.}(1985)\citenamefont {Raymer},
  \citenamefont {Walmsley}, \citenamefont {Mostowski},\ and\ \citenamefont
  {Sobolewska}}]{Raymer1985}%
  \BibitemOpen
  \bibfield  {author} {\bibinfo {author} {\bibfnamefont {M.~G.}\ \bibnamefont
  {Raymer}}, \bibinfo {author} {\bibfnamefont {I.~A.}\ \bibnamefont
  {Walmsley}}, \bibinfo {author} {\bibfnamefont {J.}~\bibnamefont
  {Mostowski}},\ and\ \bibinfo {author} {\bibfnamefont {B.}~\bibnamefont
  {Sobolewska}},\ }\bibfield  {title} {\bibinfo {title} {Quantum theory of
  spatial and temporal coherence properties of stimulated {Raman} scattering},\
  }\href {https://doi.org/10.1103/PhysRevA.32.332} {\bibfield  {journal}
  {\bibinfo  {journal} {Phys. Rev. A}\ }\textbf {\bibinfo {volume} {32}},\
  \bibinfo {pages} {332} (\bibinfo {year} {1985})}\BibitemShut {NoStop}%
\bibitem [{\citenamefont {Braunstein}\ and\ \citenamefont
  {Caves}(1994)}]{BraunsteinPRL1994}%
  \BibitemOpen
  \bibfield  {author} {\bibinfo {author} {\bibfnamefont {S.~L.}\ \bibnamefont
  {Braunstein}}\ and\ \bibinfo {author} {\bibfnamefont {C.~M.}\ \bibnamefont
  {Caves}},\ }\bibfield  {title} {\bibinfo {title} {Statistical distance and
  the geometry of quantum states},\ }\href
  {https://doi.org/10.1103/PhysRevLett.72.3439} {\bibfield  {journal} {\bibinfo
   {journal} {Phys. Rev. Lett.}\ }\textbf {\bibinfo {volume} {72}},\ \bibinfo
  {pages} {3439} (\bibinfo {year} {1994})}\BibitemShut {NoStop}%
\bibitem [{\citenamefont {Gessner}\ \emph {et~al.}()\citenamefont {Gessner},
  \citenamefont {Smerzi},\ and\ \citenamefont
  {Pezzè}}]{gessner_metrological_2019}%
  \BibitemOpen
  \bibfield  {author} {\bibinfo {author} {\bibfnamefont {M.}~\bibnamefont
  {Gessner}}, \bibinfo {author} {\bibfnamefont {A.}~\bibnamefont {Smerzi}},\
  and\ \bibinfo {author} {\bibfnamefont {L.}~\bibnamefont {Pezzè}},\
  }\bibfield  {title} {\bibinfo {title} {Metrological nonlinear squeezing
  parameter},\ }\href {https://doi.org/10.1103/PhysRevLett.122.090503}
  {\bibfield  {journal} {\bibinfo  {journal} {Phys. Rev. Lett.}\ }\textbf
  {\bibinfo {volume} {122}},\ \bibinfo {pages} {090503}}\BibitemShut {NoStop}%
\bibitem [{\citenamefont {Seveso}\ \emph {et~al.}(2019)\citenamefont {Seveso},
  \citenamefont {Albarelli}, \citenamefont {Genoni},\ and\ \citenamefont
  {Paris}}]{Seveso_2020}%
  \BibitemOpen
  \bibfield  {author} {\bibinfo {author} {\bibfnamefont {L.}~\bibnamefont
  {Seveso}}, \bibinfo {author} {\bibfnamefont {F.}~\bibnamefont {Albarelli}},
  \bibinfo {author} {\bibfnamefont {M.~G.}\ \bibnamefont {Genoni}},\ and\
  \bibinfo {author} {\bibfnamefont {M.~G.~A.}\ \bibnamefont {Paris}},\
  }\bibfield  {title} {\bibinfo {title} {On the discontinuity of the quantum
  fisher information for quantum statistical models with parameter dependent
  rank},\ }\href {https://doi.org/10.1088/1751-8121/ab599b} {\bibfield
  {journal} {\bibinfo  {journal} {Journal of Physics A: Mathematical and
  Theoretical}\ }\textbf {\bibinfo {volume} {53}},\ \bibinfo {pages} {02LT01}
  (\bibinfo {year} {2019})}\BibitemShut {NoStop}%
\bibitem [{Note1()}]{Note1}%
  \BibitemOpen
  \bibinfo {note} {This target is motivated by potential photodamage or the
  molecules. Depending on experimental considerations, other optimization
  targets may also be imagined.}\BibitemShut {Stop}%
\bibitem [{Note2()}]{Note2}%
  \BibitemOpen
  \bibinfo {note} {We clarify here our model is insufficient to describe the
  spontaneous regime adequately, since only two modes are included in Eq.~(\ref
  {eq.K_SRS}). The 'true' spontaneous emission is isotropic, potentially
  necessitating a multimode description.}\BibitemShut {Stop}%
\bibitem [{\citenamefont {Schlawin}\ and\ \citenamefont
  {Buchleitner}(2017)}]{Schlawin2017}%
  \BibitemOpen
  \bibfield  {author} {\bibinfo {author} {\bibfnamefont {F.}~\bibnamefont
  {Schlawin}}\ and\ \bibinfo {author} {\bibfnamefont {A.}~\bibnamefont
  {Buchleitner}},\ }\bibfield  {title} {\bibinfo {title} {Theory of coherent
  control with quantum light},\ }\href
  {https://doi.org/10.1088/1367-2630/aa55ec} {\bibfield  {journal} {\bibinfo
  {journal} {New Journal of Physics}\ }\textbf {\bibinfo {volume} {19}},\
  \bibinfo {pages} {013009} (\bibinfo {year} {2017})}\BibitemShut {NoStop}%
\bibitem [{\citenamefont {Gessner}\ \emph {et~al.}(2018)\citenamefont
  {Gessner}, \citenamefont {Pezz\`e},\ and\ \citenamefont
  {Smerzi}}]{PhysRevLett.121.130503}%
  \BibitemOpen
  \bibfield  {author} {\bibinfo {author} {\bibfnamefont {M.}~\bibnamefont
  {Gessner}}, \bibinfo {author} {\bibfnamefont {L.}~\bibnamefont {Pezz\`e}},\
  and\ \bibinfo {author} {\bibfnamefont {A.}~\bibnamefont {Smerzi}},\
  }\bibfield  {title} {\bibinfo {title} {Sensitivity bounds for multiparameter
  quantum metrology},\ }\href {https://doi.org/10.1103/PhysRevLett.121.130503}
  {\bibfield  {journal} {\bibinfo  {journal} {Phys. Rev. Lett.}\ }\textbf
  {\bibinfo {volume} {121}},\ \bibinfo {pages} {130503} (\bibinfo {year}
  {2018})}\BibitemShut {NoStop}%
\bibitem [{\citenamefont {Liu}\ \emph {et~al.}(2019)\citenamefont {Liu},
  \citenamefont {Yuan}, \citenamefont {Lu},\ and\ \citenamefont
  {Wang}}]{Liu_2020}%
  \BibitemOpen
  \bibfield  {author} {\bibinfo {author} {\bibfnamefont {J.}~\bibnamefont
  {Liu}}, \bibinfo {author} {\bibfnamefont {H.}~\bibnamefont {Yuan}}, \bibinfo
  {author} {\bibfnamefont {X.-M.}\ \bibnamefont {Lu}},\ and\ \bibinfo {author}
  {\bibfnamefont {X.}~\bibnamefont {Wang}},\ }\bibfield  {title} {\bibinfo
  {title} {Quantum fisher information matrix and multiparameter estimation},\
  }\href {https://doi.org/10.1088/1751-8121/ab5d4d} {\bibfield  {journal}
  {\bibinfo  {journal} {Journal of Physics A: Mathematical and Theoretical}\
  }\textbf {\bibinfo {volume} {53}},\ \bibinfo {pages} {023001} (\bibinfo
  {year} {2019})}\BibitemShut {NoStop}%
\bibitem [{\citenamefont {Albarelli}\ \emph {et~al.}(2020)\citenamefont
  {Albarelli}, \citenamefont {Barbieri}, \citenamefont {Genoni},\ and\
  \citenamefont {Gianani}}]{ALBARELLI2020126311}%
  \BibitemOpen
  \bibfield  {author} {\bibinfo {author} {\bibfnamefont {F.}~\bibnamefont
  {Albarelli}}, \bibinfo {author} {\bibfnamefont {M.}~\bibnamefont {Barbieri}},
  \bibinfo {author} {\bibfnamefont {M.}~\bibnamefont {Genoni}},\ and\ \bibinfo
  {author} {\bibfnamefont {I.}~\bibnamefont {Gianani}},\ }\bibfield  {title}
  {\bibinfo {title} {A perspective on multiparameter quantum metrology: From
  theoretical tools to applications in quantum imaging},\ }\href
  {https://doi.org/https://doi.org/10.1016/j.physleta.2020.126311} {\bibfield
  {journal} {\bibinfo  {journal} {Physics Letters A}\ }\textbf {\bibinfo
  {volume} {384}},\ \bibinfo {pages} {126311} (\bibinfo {year}
  {2020})}\BibitemShut {NoStop}%
\bibitem [{\citenamefont {Gessner}\ \emph {et~al.}(2020)\citenamefont
  {Gessner}, \citenamefont {Smerzi},\ and\ \citenamefont
  {Pezzè}}]{Gessner2020}%
  \BibitemOpen
  \bibfield  {author} {\bibinfo {author} {\bibfnamefont {M.}~\bibnamefont
  {Gessner}}, \bibinfo {author} {\bibfnamefont {A.}~\bibnamefont {Smerzi}},\
  and\ \bibinfo {author} {\bibfnamefont {L.}~\bibnamefont {Pezzè}},\
  }\bibfield  {title} {\bibinfo {title} {Multiparameter squeezing for optimal
  quantum enhancements in sensor networks},\ }\href
  {https://doi.org/10.1038/s41467-020-17471-3} {\bibfield  {journal} {\bibinfo
  {journal} {Nature Communications}\ }\textbf {\bibinfo {volume} {11}},\
  \bibinfo {pages} {3817} (\bibinfo {year} {2020})}\BibitemShut {NoStop}%
\end{thebibliography}%

\end{document}